\documentclass{jfm}
\usepackage{graphicx}
\usepackage{epstopdf, epsfig}
\usepackage{amsmath}
\usepackage{mathtools}
\usepackage{subfigure}
\usepackage{booktabs}
\usepackage{tikz}
\usepackage{multirow}
\usepackage{pstool}
\usepackage{psfrag}
\usepackage{pictex}
\usepackage{pifont}
\usepackage{xcolor}
\usetikzlibrary{shapes}
\usetikzlibrary{shapes.misc}
\usepackage{float}

\shorttitle{Resolvent-based feedback control}
\shortauthor{B. Jin, S. J. Illingworth and R. D. Sandberg}

\title{Feedback control of vortex shedding using a resolvent-based modelling approach}
\author{Bo Jin
  \corresp{\email{bjin1@student.unimelb.edu.au}},
  Simon J.  Illingworth
 \and Richard D. Sandberg}

\affiliation{Department of Mechanical Engineering, University of Melbourne,
VIC, 3010, Australia}

\DeclarePairedDelimiter\norm{\lVert}{\rVert}

\newcommand{\xmark}{}%
\DeclareRobustCommand{\xmark}{%
	\tikz\draw[scale=0.5, color=black,line width = 1pt]
	(0,.2) -- (.2,0) (0.0,0.0) -- (0.2,0.2) -- cycle;%
}

\newcommand{\blackline}{\raisebox{2pt}{\tikz{\draw[black,solid,line width = 1pt](0,0) -- (5mm,0);}}}
\newcommand{\blacklineshort}{\raisebox{2pt}{\tikz{\draw[black,solid,line width = 1pt](0,0) -- (1.5mm,0);}}}
\newcommand{\blackdot}{\tikz{\draw[black,fill=black] (0,0) circle (.4ex)}}

\newcommand{\bluelineshort}{\raisebox{2pt}{\tikz{\draw[blue,solid,line width = 1pt](0,0) -- (1.5mm,0);}}}

\newcommand{\bluedot}{\tikz{\draw[blue,fill=blue] (0,0) circle (.4ex)}}
\newcommand{\bluesmalldot}{\raisebox{1.5pt}{\tikz{\draw[blue,fill=blue] (0,0) circle (.2ex);}}}

\newcommand{\reddot}{\tikz{\draw[red,fill=red] (0,0) circle (.4ex)}}
\newcommand{\redlineshort}{\raisebox{2pt}{\tikz{\draw[red,solid,line width = 1pt](0,0) -- (1.5mm,0);}}}
\newcommand{\redcir}{\tikz{\draw[red,fill=white,line width = 1pt] (0,0) circle (.35ex)}}

\newcommand{\mytriangle}[1]{\tikz{\node[draw=#1,fill=#1,isosceles
triangle,isosceles triangle stretches,shape border rotate=90,minimum
width=0.2cm,minimum height=0.2cm,inner sep=0pt] at (0,0) {};}}

\begin{document}

\maketitle

\begin{abstract}
An investigation of optimal feedback controllers' performance and robustness is carried out for vortex shedding behind a 2D cylinder at low Reynolds numbers. To facilitate controller design, we present an efficient modelling approach in which we utilise the resolvent operator to recast the linearised Navier-Stokes equations into an input-output form from which frequency responses can be computed. The difficulty of applying modern control design techniques to complex, high-dimensional flow systems is thus overcome by using low-order models identified from these frequency responses. The low-order models are used to design optimal control laws using $\mathcal{H}_{\infty}$ loop shaping. Two distinct control arrangements are considered, both of which employ a single-input and a single-output. In the first control arrangement, a velocity sensor located in the wake drives a pair of body forces near the cylinder. Complete suppression of shedding is observed up to a Reynolds number of $\Rey=110$. Due to the convective nature of vortex shedding and the corresponding time delays, we observe a fundamental trade-off: the sensor should be close enough to the cylinder to avoid any excessive time lag, but it should be kept sufficiently far from the cylinder to measure any unstable modes developing downstream. It is found that these two conflicting requirements become more difficult to satisfy for larger Reynolds numbers. In the second control arrangement, we consider a practical setup with a body-mounted force sensor and an actuator that oscillates the cylinder according to the lift measurement. It is shown that the system is stabilised only up to $\Rey=100$, and we demonstrate why the performance of the resulting feedback controllers deteriorates much more rapidly with increasing Reynolds number. The challenges of designing robust controllers for each control setup are also analysed and discussed. 
\end{abstract}

\section{Introduction}

In the last decades, the control of fluid dynamics has received considerable attention from researchers and engineers. Indeed, the ability to manipulate fluid flows can be of tremendous benefit in a number of applications. A typical example is the flow past a circular cylinder which becomes unstable above the critical Reynolds number near $47$ and results in vortex-shedding \citep{jackson1987finite,provansal1987benard,zebib1987stability,duvsek1994numerical}. This gives rise to strong force fluctuations which are responsible for structural vibrations, acoustic noise and resonance \citep{williamson1996vortex}. Much research has been conducted on the suppression of vortex shedding using either passive strategies (without additional energy input) or active strategies (with an external energy source), as summarised by \citet{choi2008control}.

\subsection{Feedback control and model reduction}
Closed-loop control, where actuators operate actively according to real-time measurements of the flow field, is a robust and energetically efficient way to control fluid flows. In some studies, feedback control has been applied without knowledge of the flow dynamics. Simple proportional feedback laws were designed by trial and error to eliminate vortex shedding in various control arrangements, such as loudspeakers \citep{roussopoulos1993feedback}, blowing$/$suction slots \citep{park1994feedback,gunzburger1996feedback} and small rotary control cylinders \citep{muddada2010active}. \citet{zhang2004closed} developed a proportional-integral-derivative controller to suppress vortex-induced vibration on a spring-supported cylinder. A further investigation of proportional, proportional-derivative, proportional-integral and proportional-integral-derivative feedback control was carried out \citep[see][]{son2011proportional,son2018iterative} to compare and optimize control laws for vortex shedding suppression. The parameters in these control laws were chosen based on either a brute-force approach or physical intuition. 

Although these studies have met with some success, they are not model-based. There exists a large set of well developed and powerful control design tools providing efficient and robust methods for model-based control design. However, due to the non-linearity and high-dimensionality of the Navier--Stokes equations, they are in general computationally intractable to apply directly to fluid flows. This difficulty can be bypassed by means of flow linearisation \citep{kim2007linear,sipp2010dynamics} and model reduction \citep{taira2017modal,rowley2017model}, where important features of fluid flows are approximated by low-order linear models. 

A number of techniques have been developed to approximate flow dynamics by reduced-order models (ROMs) for the purpose of control design, such as POD (proper orthogonal decomposition), balanced POD and the ERA (Eigensystem Realisation Algorithm). One of the earliest studies is that of \citet{aubry1988dynamics}, who approximated important features of the boundary layer by a low-dimensional nonlinear model based on POD. The method has shown its effectiveness in later research for flow control problems, such as the cylinder wake \citep{gillies1998low,singh2001optimal} and channel flow \citep{ilak2008modeling}. 

Much recent work in flow control has focused on linear input-output formulations of flow systems \citep{bagheri2009input}. Using this approach, the governing equations are projected onto the most controllable and observable modes to capture important input-output dynamics. These modes can be generated by either balanced POD \citep{rowley2005model} or the ERA \citep{juang1985eigensystem}. A detailed comparison between these methods was presented by \citet{ma2011reduced} using the example of the flow past an inclined flat plate. Without using any adjoint information, ERA produced the same ROMs as those given by balanced POD, which allows its direct application to flow systems using only simulation or experimental data \citep{belson2013feedback,illingworth2016model,flinois2016feedback,yao2017feedback,yao2017model}. Recently, model reduction based on resolvent analysis has shown good potential. The method has been applied to efficiently identify physical flow structures in a broad range of nonlinear flows, such as cavity flow \citep{gomez2016reduced}, the flat-plate boundary layer \citep{sipp2013characterization}, pipe flow \citep{mckeon2010critical} and cylinder flow \citep{symon2018non}. 

Some recent studies have demonstrated a significant influence of actuator and sensor choice and placement on the performance and robustness of feedback controllers. Studies of the one-dimensional complex Ginzburg-Landau system indicated that the optimal actuator and sensor placement could significantly influence the performance of feedback controllers \citep{chen2011h,oehler2018sensor}. In a two-dimensional Blasius boundary layer, the type and relative position of the sensor and the actuator were also shown to be crucial to the controller's properties \citep{belson2013feedback}.

\subsection{This article}
The current work uses an efficient modelling approach with an input-output formulation based on resolvent analysis \citep{sipp2013characterization}. This allows one to directly obtain frequency responses for a broad range of control configurations. The MATLAB package VECTFIT \citep{gustavsen2013vector} is used to fit reduced-order models to these frequency responses. The $\mathcal{H}_{\infty}$ loop shaping method \citep{glover1989robust} is employed to design optimal feedback controllers and also provides a stability margin which serves as an indicator of the control performance achieved. We demonstrate the effectiveness of the method for the feedback control of the instabilities that lead to vortex shedding for the flow past a 2D circular cylinder.

Another contribution of this work is to analyse the effect of different control configurations on control performance and robustness. We shall see that in-flow actuator and sensor outperforms the body-mounted setup. A further investigation about optimal in-flow sensor placement in a 2D flow allows us to observe a fundamental trade-off which is consistent with the conclusions of \citet{belson2013feedback} and \citet{oehler2018sensor}. We show that the deterioration of the control performance in a bluff body flow also has a variety of root causes \citep{hoagg2007nonminimum}. By examining different sensor positions for a range of Reynolds numbers, we demonstrate how these roots affect the performance of controllers as well as possible physical mechanisms behind them.

The article starts with the definition of the flow and control configurations in $\S$ \ref{sec:pro_formu}. In $\S$ \ref{sec:methods}, we form reduced-order models using an input-output framework and explain the numerical setups and control design techniques. The results and analysis of the in-flow and body-mounted control setups are shown in $\S$ \ref{sec:in-flow_case} and $\S$ \ref{sec:bodyset} respectively.  Conclusions are given in $\S$ \ref{sec:cons}.
\section{Problem formulation}\label{sec:pro_formu}
The objective of feedback control is to completely suppress vortex shedding behind a two-dimensional circular cylinder. In other words, we attempt to drive the flow towards its unstable steady state (base flow), around which the Navier--Stokes equations are linearised. The governing equations are the incompressible Navier--Stokes equations with external forcing: 
\begin{equation}\label{equ:nsequation}
\frac{\partial\textbf{\textit{u}}}{\partial t}+\textbf{\textit{u}}\cdot\nabla\textbf{\textit{u}}=-\nabla \textit{p}+\nu \nabla^2\textbf{\textit{u}}+\textbf{\textit{f}}',\ \ \nabla\cdot\textbf{\textit{u}}=0\ ,
\end{equation}
where the source term $\textbf{\textit{f}}'=\left[{\textit{f}_x}'\ \ {\textit{f}_y}'\right]^T$ in the momentum equation models the external forcing which is assumed to have zero mean. To investigate the instability of the cylinder flow near its steady state, we perform an input-output analysis by linearising about the laminar base flow
$(\textbf{\textit{U}},\textit{P})$:
\begin{equation}\label{equ:pertur}
\frac{\partial\textbf{\textit{u}}'}{\partial t}+ \textbf{\textit{U}}\cdot\nabla\textbf{\textit{u}}'+\textbf{\textit{u}}'\cdot\nabla\textbf{\textit{U}}=-\nabla\textit{p}'+\nu\nabla^2\textbf{\textit{u}}'+\textbf{\textit{f}}',\ \ \nabla\cdot\textbf{\textit{u}}'=0\ .
\end{equation}
Here, $(\textbf{\textit{u}}',\ \textit{p}')$ represents the unsteady components of velocity and pressure which are assumed to be small perturbations about the steady state, and $\nu$ is the kinematic viscosity. The Reynolds number is defined as $\Rey={\textit{U}_{\infty}D}/{\nu}$ where $\textit{U}_{\infty}$ is the free-stream velocity, and $D$ is the diameter of the cylinder. The range of Reynolds numbers considered is $\Rey \in [50,\ 110]$ and all lengths are non-dimensionalised by $\textit{D}$. At these Reynolds numbers the cylinder wake has a single linearly unstable mode which drives the flow to periodic self-sustained limit-cycle oscillations (vortex shedding). The controllers we design aim to keep perturbations small, and so the nonlinear term $\textbf{\textit{u}}'\cdot\nabla\textbf{\textit{u}}'$ can be neglected. This formulation allows us to use existing linear control theory and analysis techniques.

In-flow and body-mounted sensors and actuators are used and compared. We choose two simple setups as representatives of in-flow and body-mounted control arrangements and figure \ref{fig:controlsetups} shows a schematic of them.
\subsection{In-flow control setup}
In the first control setup, the momentum equation \eqref{equ:pertur} is forced by a pair of anti-symmetrical body forces $\textbf{\textit{f}}'$ that serve as an in-flow actuator:
\begin{equation}
    \textbf{\textit{f}}'=\textbf{B}(x,y)\left[\textit{u}(\textit{t})+\textit{w}(\textit{t})\right]\ ,
\end{equation}
where $\textit{u}(\textit{t})$ is the actuator signal provided by a transverse velocity sensor positioned a distance $d$ downstream of the cylinder, and $\textit{w}(\textit{t})$ is a disturbance from the actuator. $\textbf{B}(x,y)$ is the spatial distribution of the actuator and disturbance. More specifically, 
\begin{equation}
    \textbf{B}=
    \begin{bmatrix}
    \cos{\theta}\left(\mathcal{S}(A,\sigma,r,\theta)-\mathcal{S}(A,\sigma,r,-\theta)\right)\\
    \sin{\theta}\left(\mathcal{S}(A,\sigma,r,\theta)+\mathcal{S}(A,\sigma,r,-\theta)\right)
    \end{bmatrix}\ ,\ \textrm{where}
\end{equation}
\begin{equation}
    \mathcal{S}(A,\sigma,r,\theta)=\frac{A}{2\pi\sigma^2}exp\left(-\frac{\left(x-r\cos{\theta}\right)^2+\left(y-r\sin{\theta}\right)^2}{2\sigma^2}\right)\ ,
\end{equation}
where the magnitude is $A=1.0$ and the standard deviation $\sigma=0.1$. The centres of the above distributions are near the separation points at a distance of $r=0.6$ from the cylinder's centre and an angle of $\theta=\pm70^{\circ}$ from the cylinder's downstream-pointing horizontal. We fix the form and the location of the actuator and vary only the sensor position along the centreline to investigate the influence of sensor placement. This arrangement is similar to that used by \citet{illingworth2016model}.
\subsection{Body-mounted control setup}
The second control setup considered uses a body-mounted actuator and a body-mounted sensor. Similar to a fluid-structure interaction (FSI) system, we consider an oscillatory cylinder whose transverse acceleration $\textit{a}$ is controlled according to the feedback signal provided by a lift sensor attached on the cylinder. Optimal feedback controllers are designed for different Reynolds numbers to suppress vortex shedding and the vibration of the bluff body.
\begin{figure}
	\centering
		\includegraphics[width=0.95\textwidth]{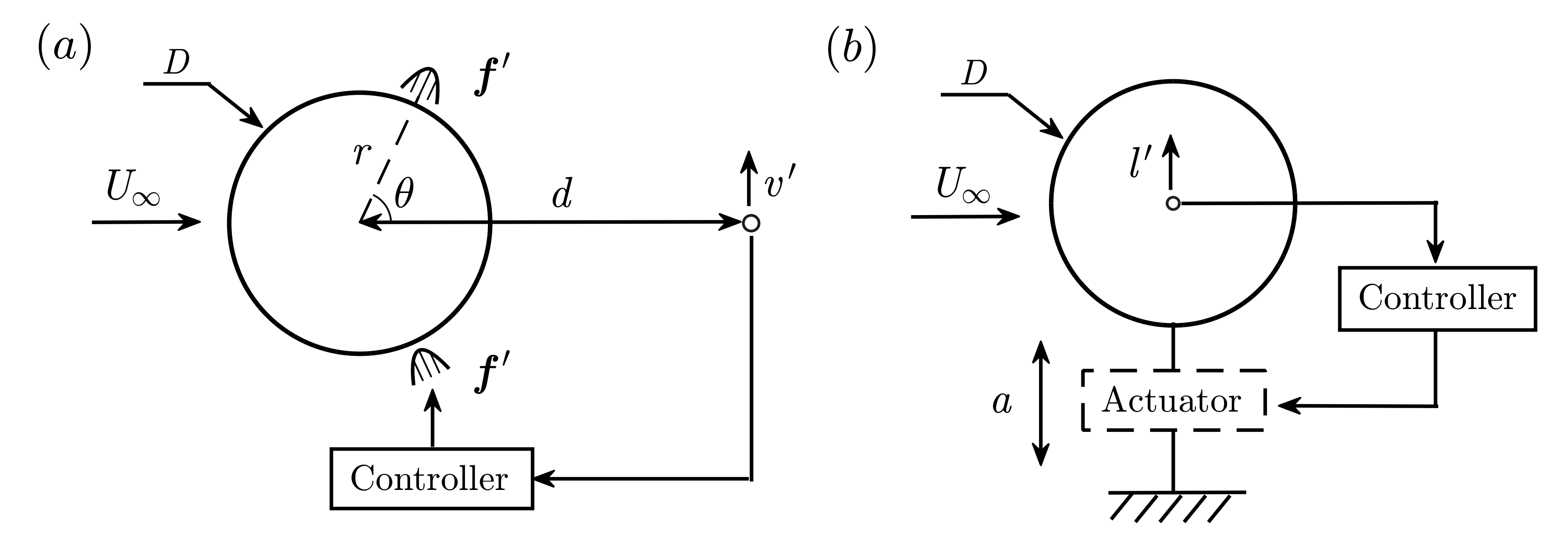}
	\caption{$(\textit{a})$ A pair of body forces $\textbf{\textit{f}}'$ applied near the cylinder surface cooperates with a velocity sensor placed on the centreline with a distance $\textit{d}$ from the centre of the cylinder.  $(\textit{b})$ The force sensor is placed on the cylinder to measure the lift $\textit{l}\hspace{0.25mm}'$ which is fed to the actuator that controls the transverse acceleration $\textit{a}$ of the cylinder.}
	\label{fig:controlsetups}
\end{figure}

\section{Modelling and control methods}\label{sec:methods}
\subsection{State-space formulation and system identification}
The linearised Navier-Stokes equations can be written in standard state-space form which is useful for performing input-output analysis. To evaluate the linear dynamics of a flow system, it is convenient to take Laplace transforms. Introducing the transformation into equation \eqref{equ:pertur}, we obtain the equations
\begin{equation}\label{equ:source}
s\hat{\textbf{\textit{u}}}+ \textbf{\textit{U}}\cdot\nabla\hat{\textbf{\textit{u}}}+\hat{\textbf{\textit{u}}}\cdot\nabla\textbf{\textit{U}}=-\nabla\hat{\textit{p}}+\nu\nabla^2\hat{\textbf{\textit{u}}}+\hat{\textbf{\textit{f}}},\ \ \nabla\cdot\hat{\textbf{\textit{u}}}=0\ ,
\end{equation}
where $(\hat{\textbf{\textit{u}}},\ \hat{\textit{p}})$ and $\hat{\textbf{\textit{f}}}$ represent the (complex-valued) spatial structure of the velocity and body forcing and $s=\sigma+\textit{j}\omega$ is the Laplace variable. Thus, the transfer function between the external forcing and response can be written as
\begin{equation}\label{equ:pertur_sys}
\begin{bmatrix}
\hat{\textbf{\textit{u}}}\\
\hat{\textit{p}}
\end{bmatrix}=
(s\mathcal{E}-\mathcal{A})^{-1}
\begin{bmatrix}
\hat{\textbf{\textit{f}}}\\
0
\end{bmatrix}\ ,
\end{equation}
where $(s\mathcal{E}-\mathcal{A})^{-1}$ is known as the resolvent operator and $\mathcal{A}$ is the linearised Navier--Stokes operator around the base flow:
\begin{equation}\label{equ:pertur_matrix}
\mathcal{A}=
\begin{bmatrix}
-\textbf{\textit{U}}\cdot\nabla-()\cdot\nabla\textbf{\textit{U}}+\nu\nabla^2&-\nabla\\
\nabla\cdot()&0
\end{bmatrix},\ \ 
\mathcal{E}=
\begin{bmatrix}
\textit{I}&0\\
0&0
\end{bmatrix}\ .
\end{equation}
We now consider writing equation \eqref{equ:pertur_sys} in state-space form for a linear, time-variant dynamical system $P(s)$:
\begin{eqnarray}\label{equ:state-space}
s\mathcal{E}\textbf{\textit{x}}&=&\mathcal{A}\textbf{\textit{x}}+\mathcal{B}\textit{e}\nonumber\\
\textit{y}&=&\mathcal{C}\textbf{\textit{x}}+\mathcal{D}\textit{e}\ ,
\end{eqnarray}
where $\textbf{\textit{x}}=\left[\hat{\textbf{\textit{u}}}\ \ \hat{\textit{p}}\right]^T$ is the system state, $\textit{e}$ is an input vector of dimension $p$ and $\textit{y}$ is an output vector of dimension $q$. The feedback control arrangement used here is single-input single-output (SISO) with $p=q=1$. The vector $\mathcal{B}$ is determined by the shape of the actuation, which is from the spatial discretisation of the external forcing term  $\hat{\textbf{\textit{f}}}$. The matrices $\mathcal{C}$ and $\mathcal{D}$ represent output and feed-forward dynamics respectively. The transfer function between the output $\textit{y}$ and input $\textit{e}$ can then be written as
\begin{equation}\label{equ:modelP}
P(s)=\mathcal{C}\left(s\mathcal{E}-\mathcal{A}\right)^{-1}\mathcal{B}+\mathcal{D}\ .
\end{equation}

In general, $P(s)$ is of high dimension, which makes the control design problem computationally intractable. However, it is feasible to solve this linear system and get the frequency response data from the actuation to the measurement. Reduced order models can then be formed for the corresponding frequency responses using a vector-fitting algorithm VECTFIT \citep{gustavsen1999rational,gustavsen2006improving,deschrijver2008macromodeling}. VECTFIT identifies a reduced-order model $\widetilde{P}(s)$ of a significantly smaller dimension $N$ in the rational form:
\begin{equation}\label{equ: modelPtilda}
    \widetilde{P}(s)=\sum_{m=1}^{N}\dfrac{r_m}{s-a_m}+sh+d\ ,
\end{equation}
where the poles $a_m$, residues $r_m$, and terms $d$ and $h$ are identified such that the frequency response of $\widetilde{P}(s)$ is within a distance $\epsilon$ of the original system $P(s)$ over a broad frequency range.
\begin{figure}
  \centerline{\includegraphics[width=0.5\textwidth]{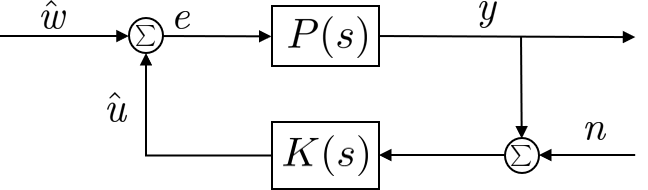}}
  \caption{Block diagram of the closed-loop control system.}
\label{fig:blockdia}
\end{figure}

\subsection{Feedback controller design}
The feedback controller is designed based on the reduced-order model $\widetilde{P}(s)$ using $\mathcal{H}_{\infty}$ loop-shaping. The control block diagram is shown in figure \ref{fig:blockdia} where $\hat{\textit{w}}$ represents disturbances from the actuator and $\textit{n}$ represents noise at the sensor. We assume a positive feedback configuration as illustrated in figure \ref{fig:blockdia}. The controller $K(s)$ for plant $P(s)$ is designed using the loop shaping design method of \citet{glover1989robust} which maximises the normalised coprime stability margin $\textit{b}(\widetilde{P},K)$ of the plant-controller feedback loop
\begin{equation}
    \textit{b}=
    	\norm[\bigg]{\begin{bmatrix}
    		K\\
    		I
    		\end{bmatrix}
    		\left(I+\widetilde{P}K\right)^{-1}
    		\begin{bmatrix}
    		\widetilde{P}&I
    		\end{bmatrix}
    	}^{-1}_{\infty}\ ,
\end{equation}
where $\textit{b}\in [0,1]$, which can be maximised over all stabilising $K$ to give
\begin{equation}\label{equ:bopt}
    \textit{b}_{opt}(\widetilde{P})=\sup_{K} \textit{b}(\widetilde{P},K)\ .
\end{equation}

Physically, $\textit{b}_{opt}$ is an indication of the robustness of the closed-loop system to unmodelled dynamics and also serves as a performance measure: the larger $\textit{b}_{opt}$, the better the performance and the greater the robustness of the closed-loop system.

For SISO systems as considered here, a compensator which weights the plant according to the control objectives is used of the form
\begin{equation}\label{equ:weight}
\textit{W}(s)=k\frac{a^2}{(s+a)^2}\ .
\end{equation} 

The parameters $k$ and $a$ are chosen such that the gain of the weighted plant is sufficiently high at frequencies where good disturbance attenuation is required and is sufficiently low at high frequencies where modelling uncertainties will be greatest \citep{illingworth2016model}. 
\begin{figure}
  \centerline{\includegraphics[width=0.65\textwidth]{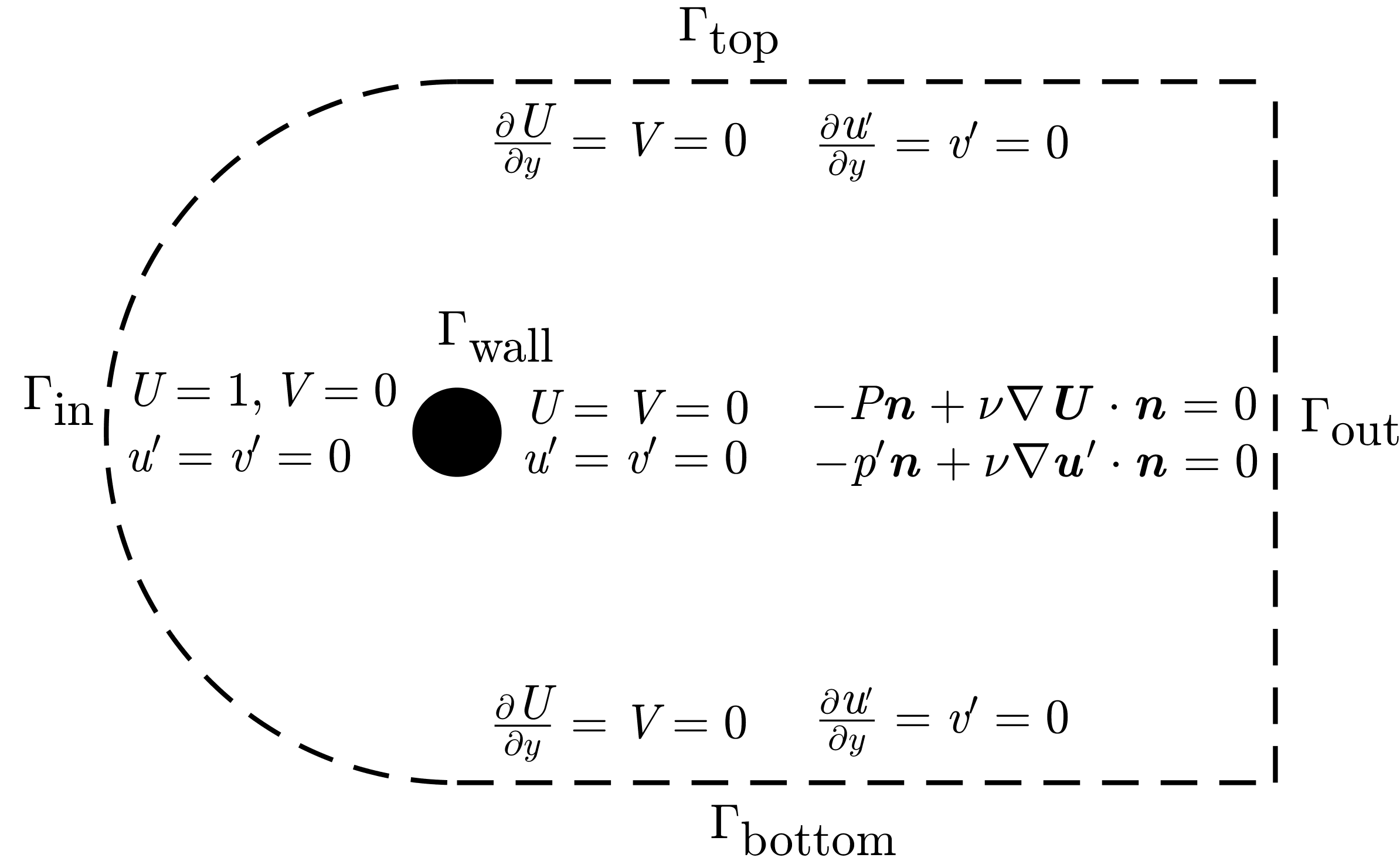}}
  \caption{Computational domain and boundary conditions for the steady and unsteady linearised Navier-Stokes equations for the in-flow setup.}
\label{fig:compudomain}
\end{figure}
\subsection{Numerical setup}
We consider an incompressible flow with free-stream velocity $\textit{U}_{\infty}$ past a two-dimensional circular cylinder of diameter $\textit{D}$. Simulations are conducted on the computing platform FEniCS \citep{logg2012automated} which has been extensively used for fluid mechanics \citep{mortensen2011fenics, nguyen2019direct, vasilyeva2019constrained}. Three versions of the flow are solved: the steady solution of \eqref{equ:nsequation} (base flow), the linearised perturbation equation \eqref{equ:state-space} in state-space form and the fully nonlinear Navier-Stokes equations. We employ the same computational domain as that used by \citet{leontini2006wake}, as shown in figure \ref{fig:compudomain}, and discretize using Taylor-Hood finite elements over a structured mesh. In order to appropriately resolve the details of the flow, the mesh points are clustered smoothly near the cylinder and in the wake. More specifically, the mesh consists of $5.46\times10^4$ triangles and the minimum wall-normal size around the cylinder is $0.01$. A backward Euler scheme is used for time discretisation ($\Delta t=0.01$) in the direct numerical simulations with the maximum Courant number below 0.6 to ensure accuracy.

The base flow $(\textbf{\textit{U}},\ \textit{P})$, which is governed by the unforced steady Navier--Stokes equations, is solved using a Newton method. The corresponding boundary conditions are summarised in figure \ref{fig:compudomain}. Dirichlet boundary conditions are imposed at the inlet $\Gamma_\textrm{in}$ $(x<0)$ and at the cylinder surface $\Gamma_\textrm{wall}$ (centred at $x=y=0$):
\begin{subequations}
\begin{eqnarray}
\textit{U}=1,\hspace{0.2cm} \textit{V}=0 \hspace{1.5cm} \textrm{on}&\hspace{0.5cm} \Gamma_\textrm{in}\ ,\\
\textit{U}=0,\hspace{0.2cm} \textit{V}=0 \hspace{1.5cm} \textrm{on}&\hspace{0.5cm} \Gamma_\textrm{wall}\ .
\end{eqnarray}
\end{subequations}

Symmetry conditions are enforced on the top and bottom boundaries $\Gamma_\textrm{top}$ and $\Gamma_\textrm{bottem}$ ($0<x<23,\ y = \pm 15$):
\begin{equation}
    \dfrac{\partial \textit{U}}{\partial y}=0,\hspace{0.2cm} \textit{V}=0 \hspace{1.5cm} \textrm{on}\hspace{0.5cm} \Gamma_\textrm{top}\cup\Gamma_\textrm{bottom}\ ,\\
\end{equation}
while the pressure and the velocity are combined into the standard outflow conditions on the outlet boundary $\Gamma_\textrm{out}$ $(x=23,\ -15<y<15)$:
\begin{equation}
    -\textit{P}\textbf{\textit{n}}+\nu\nabla\textbf{\textit{U}}\cdot\textbf{\textit{n}}=0  \hspace{1.5cm} \textrm{on}\hspace{0.5cm} \Gamma_\textrm{out}\ ,
\end{equation}
where $\textbf{\textit{n}}$ denotes the outward-pointing normal vector on the boundary. The base flow is the same for all control setups. Vorticity contours for the base flow are shown in figure \ref{fig:basepertur_1}($\textit{a}$) at $\Rey=60,\ 80,\ 100$.

The basic boundary conditions for the perturbation $(\textbf{\textit{u}}',\ \textit{p}')$ are also depicted in figure \ref{fig:compudomain}. For the in-flow control setup, the linear perturbation system has the same boundary conditions as the base flow except at the inlet where homogeneous boundary conditions $(\textit{u}'=\textit{v}'=0)$ are enforced to ensure zero perturbations at infinity. In the body-mounted control setup, the flow is solved in an accelerated frame of reference attached to the cylinder instead of moving the cylinder directly. For this, the transverse acceleration of the frame $\textit{a}$ is treated as an extra forcing term $\textbf{\textit{f}}'=[0\ \ \textit{a}]^T$. Thus, the boundary conditions at the inlet, top and bottom boundaries are modified \citep{leontini2006wake}:
\begin{subequations}
\begin{eqnarray}
    \textit{u}'=0,\hspace{0.2cm} \textit{v}'=\int_0^t\textit{a}(t)\ dt \hspace{1.5cm} \textrm{on}&\hspace{0.5cm} \Gamma_\textrm{in}\ ,\\
    \dfrac{\partial \textit{u}'}{\partial y}=0,\hspace{0.2cm} \textit{v}'=\int_0^t\textit{a}(t)\ dt \hspace{1.5cm} \textrm{on}&\hspace{0.5cm} \Gamma_\textrm{top}\cup\Gamma_\textrm{bottom}\ ,
\end{eqnarray}
\end{subequations}
where zero initial conditions are assumed. In the linear state-space model \eqref{equ:state-space}, these boundary conditions are enforced in the frequency domain by applying Laplace transforms:
\begin{subequations}
\begin{eqnarray}
    \hat{\textit{u}}=0,\hspace{0.2cm} \hat{\textit{v}}=\dfrac{\hat{\textit{a}}}{s} \hspace{1.5cm} \textrm{on}&\hspace{0.5cm} \Gamma_\textrm{in}\ ,\\
    \dfrac{\partial \hat{\textit{u}}}{\partial y}=0,\hspace{0.2cm} \hat{\textit{v}}=\dfrac{\hat{\textit{a}}}{s} \hspace{1.5cm} \textrm{on}&\hspace{0.5cm} \Gamma_\textrm{top}\cup\Gamma_\textrm{bottom}\ .
\end{eqnarray}
\end{subequations}

The stability analysis of the base flows and discretised perturbation systems have been validated using the results of \citet{barkley2006linear}. To validate the controllers, direct numerical simulations are performed using the IPCS (Incremental Pressure Correction Scheme) method which has been extensively tested in \citet{logg2012automated}. The corresponding boundary conditions are modified based on the identity $(\textbf{\textit{u}}, \textit{p})=(\textbf{\textit{U}}+\textbf{\textit{u}}',\ \textit{P}+\textit{p}')$.

\section{In-flow control setup case}\label{sec:in-flow_case}
In the first instance, an in-flow control setup is considered in which a single sensor is placed in the wake and a single actuator is applied near the cylinder (see figure \ref{fig:controlsetups}$(\textit{a})$). We first form reduced-order models for model-based control, and the performance and robustness of all controllers are characterised and compared. We vary the Reynolds number and the location of the velocity sensor to investigate the influence of sensor placement on control at different Reynolds numbers.
\subsection{Open-loop system identification}\label{sec:case1}
\begin{figure}
 \vspace{2mm}
  \centerline{\includegraphics[width=0.45\textwidth]{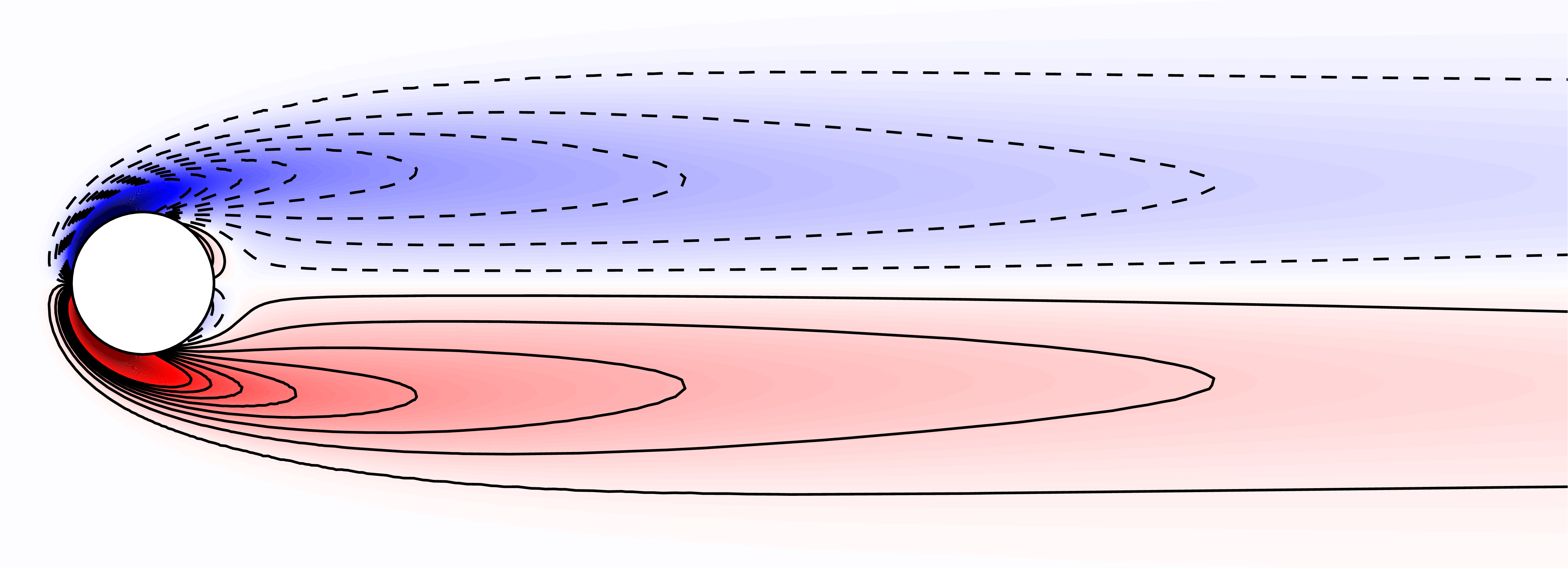}
  \llap{\parbox[b]{2.55in}{(a)\\\rule{0ex}{0.725in}}}
  \hspace{2mm}
  \includegraphics[width=0.45\textwidth]{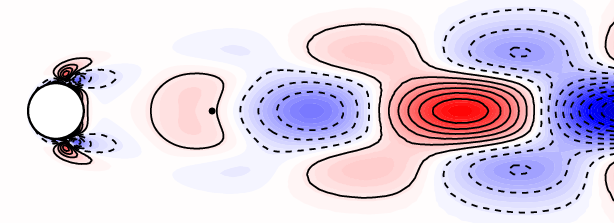}
  \llap{\parbox[b]{2.55in}{(b)\\\rule{0ex}{0.725in}}}}
    \centerline{\includegraphics[width=0.45\textwidth]{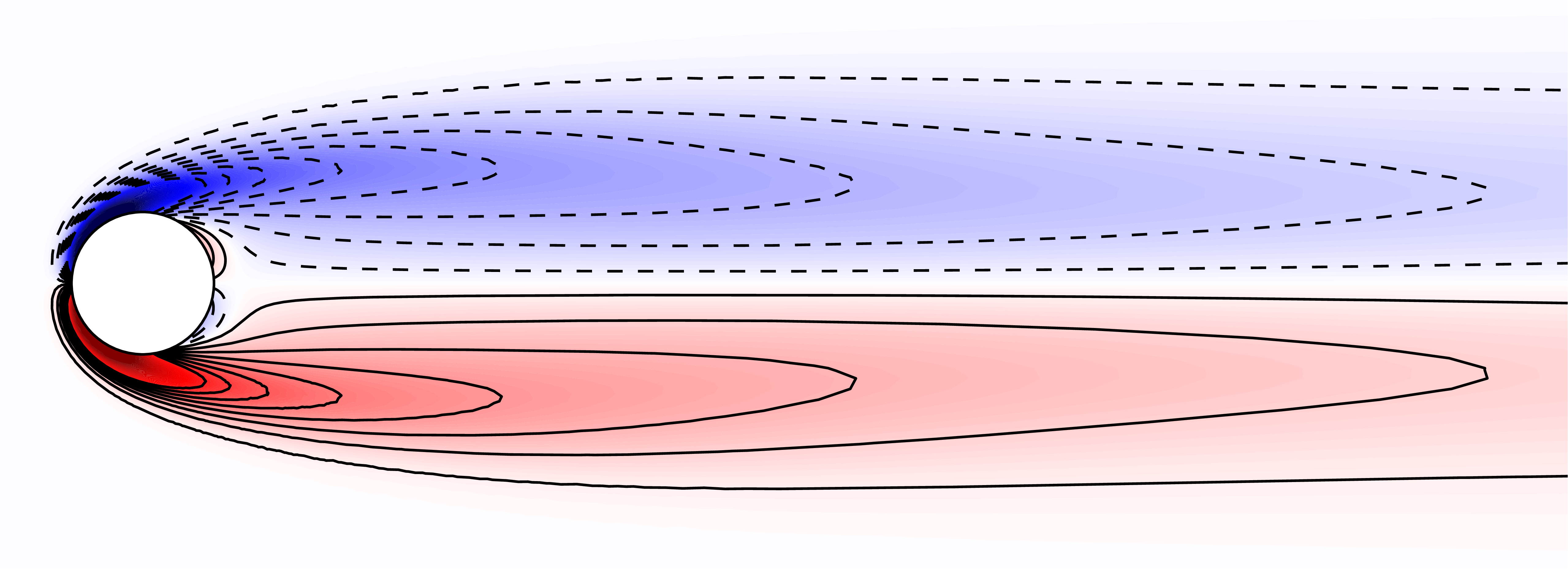}
    \llap{\parbox[b]{2.55in}{$\ $\\\rule{0ex}{0.725in}}}
    \hspace{2mm}
  \includegraphics[width=0.45\textwidth]{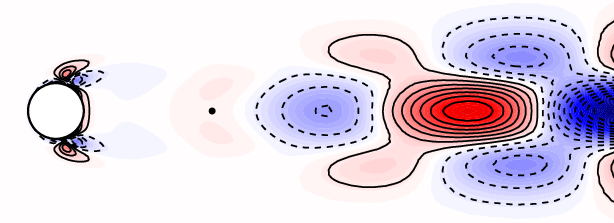}
  \llap{\parbox[b]{2.55in}{$\ $\\\rule{0ex}{0.725in}}}}
    \centerline{\includegraphics[width=0.45\textwidth]{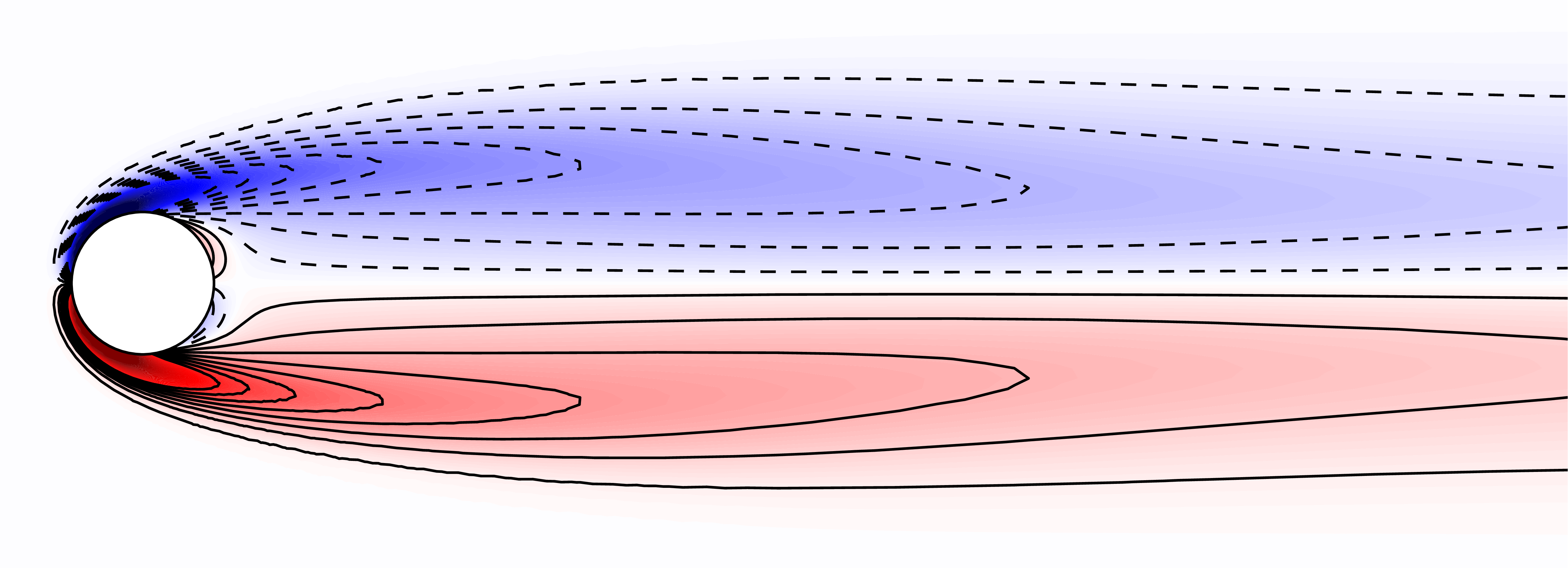}
    \llap{\parbox[b]{2.55in}{$\ $\\\rule{0ex}{0.725in}}}
    \hspace{2mm}
  \includegraphics[width=0.45\textwidth]{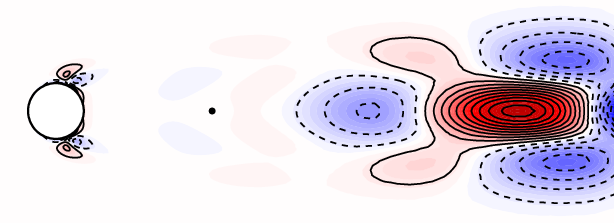}
  \llap{\parbox[b]{2.55in}{$\ $\\\rule{0ex}{0.725in}}}}
  \caption{Vorticity contours (dashed lines for negative and solid lines for positive vorticity) for (\textit{a}) the unstable base flow and (\textit{b}) the perturbation system (normalised, real part) forced by the harmonic forcing at the unstable frequency for $\Rey=60,\ 80,\ 100$ (from top to bottom). A velocity sensor (\protect\blackdot) is placed a distance $\textit{d}$ downstream the cylinder. Contours in (\textit{a})/(\textit{b}) share the same scale for comparison.}
\label{fig:basepertur_1}
\end{figure}

Figure \ref{fig:basepertur_1}($\textit{b}$) shows the normalised vorticity field of the corresponding perturbation system $P(s)$ actuated by the body force $\hat{\textbf{\textit{f}}}$ at the unstable frequency $s=j\omega_u$ (i.e. the resolvent operator between the input and output in \eqref{equ:pertur_sys}). The shear layer emanating from the separation point is disturbed by the actuator and grows into a large-scale vortical structure downstream. As the Reynolds number increases, the large vortical structure formed by the shear layer appears increasingly far downstream.

\begin{figure}
 \vspace{2mm}
 \hspace{1mm}
    \centerline{\includegraphics[width=0.95\textwidth]{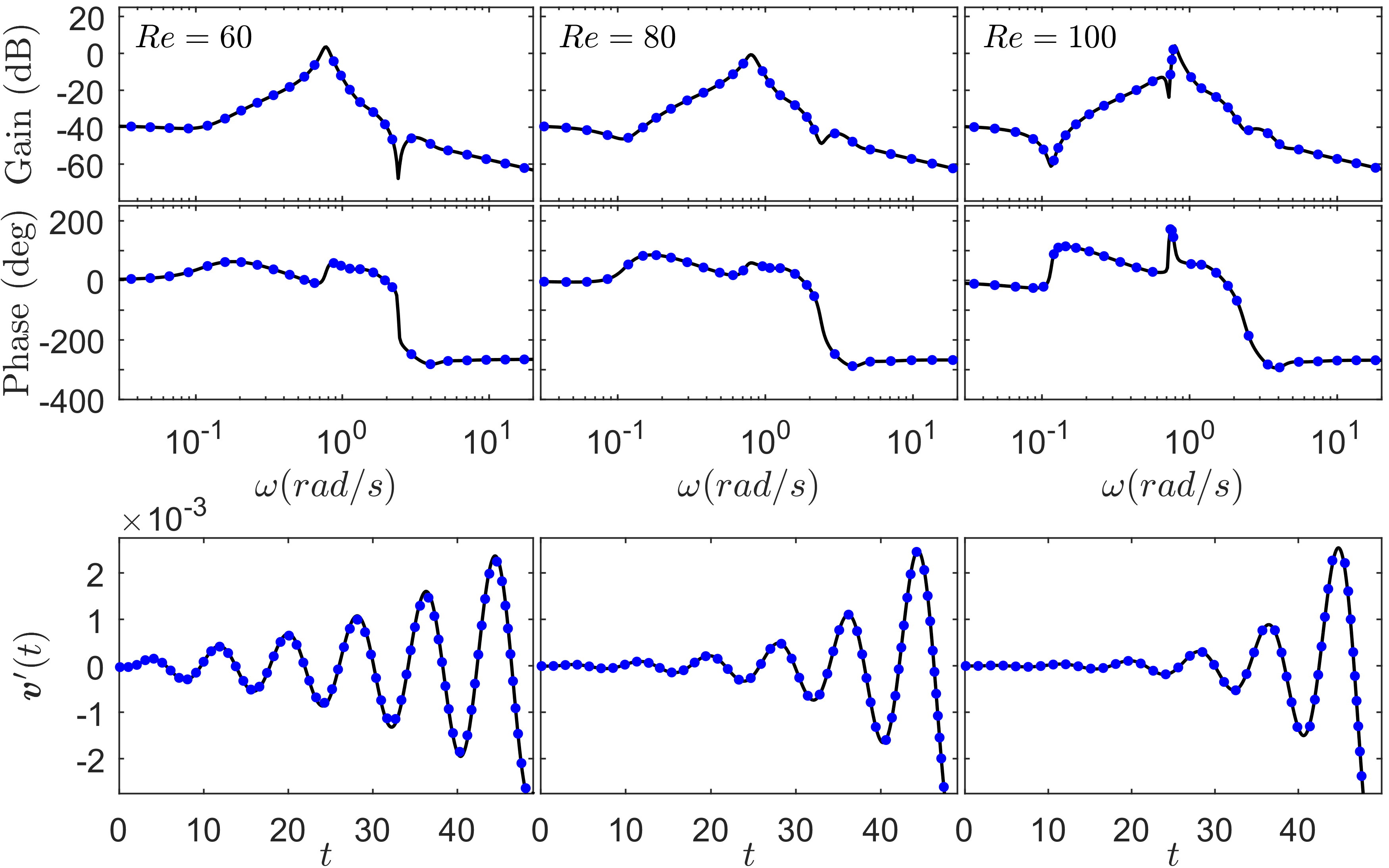}
    \llap{\parbox[b]{5.2in}{(a)\\\rule{0ex}{3.1in}}}
    \llap{\parbox[b]{5.25in}{(b)\\\rule{0ex}{1.175in}}}}
    \caption{(\textit{a}) Frequency responses from $P(j\omega)$ (\protect\blackline) compared to those from reduced-order models $\widetilde{P}(j\omega)$ (\protect\bluedot) at $\Rey=60,\ 80,\ 100$. (\textit{b}) The corresponding open-loop impulse responses from numerical simulations. The results for $\Rey=60,\ 80$ are multiplied by 20 and 3, respectively, so that the same scale can be used.}
    \label{fig:modelvalidation}
\end{figure}

After obtaining the base flow, the transfer function between the actuator signal and the sensor measurement, denoted by $\widetilde{P}(s)$, is to be determined for control design. Instead of harvesting data from direct numerical simulations, e.g. \citet{dahan2012feedback} and \citet{illingworth2016model}, we obtain frequency responses by solving \eqref{equ:modelP} directly for harmonic inputs ($s=j\omega$) over a broad range of frequencies. The reduced-order models are identified from the gain and phase of the response data by utilising a vector-fitting algorithm \citep{gustavsen2013vector}. Figure \ref{fig:modelvalidation}($\textit{a}$) shows comparisons of response data from the linear perturbation systems $P(s)$ and the identified reduced-order models $\widetilde{P}(s)$ at $\Rey=60,\ 80,\ 100$ with the velocity sensor placed at $d=2.8D$. The order (or dimension) of $\widetilde{P}(s)$ is chosen such that the fitting residual $\epsilon$ is below $10^{-5}$. This is achieved with an order of 30 or less for all Reynolds numbers considered. 

The accuracy of the reduced-order models is validated by the excellent agreement between the open-loop impulse responses from these models and direct numerical simulations (DNS), which is summarised in figure \ref{fig:modelvalidation}($\textit{b}$). Due to the small perturbation assumption, the magnitude of the impulse equals $1\times 10^{-4}$ and the linearity of the flow systems is also confirmed. Thus, the reduced-order models  found are good approximations of the true systems.
\subsection{Model-based feedback control}\label{sec:feedcon1}
Based on the identified reduced-order models, we design an optimal feedback controller for each model and ensure that the controller satisfies closed-loop stability. The parameters in the compensator weight are determined after searching different value combinations to achieve the largest optimal stability margin $\textit{b}_{opt}$. 

\begin{figure}
\vspace{1.5mm}
\hspace{1.75mm}
    \begin{minipage}[b]{0.5\textwidth}
    \centering
    \includegraphics[width=1\textwidth]{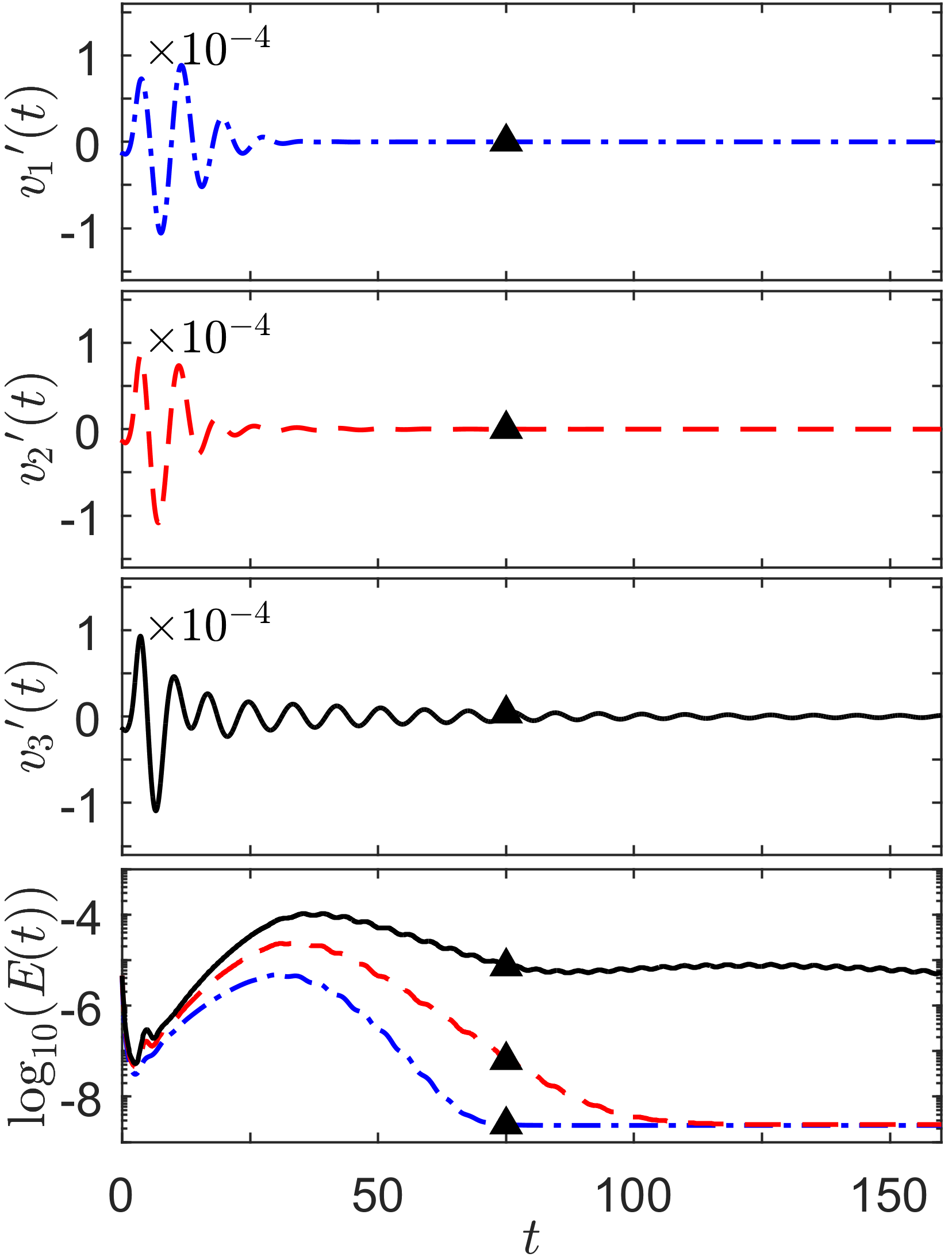}
    \end{minipage}
    \begin{minipage}[b]{0.45\textwidth}
        \centering
        \includegraphics[width=1\textwidth]{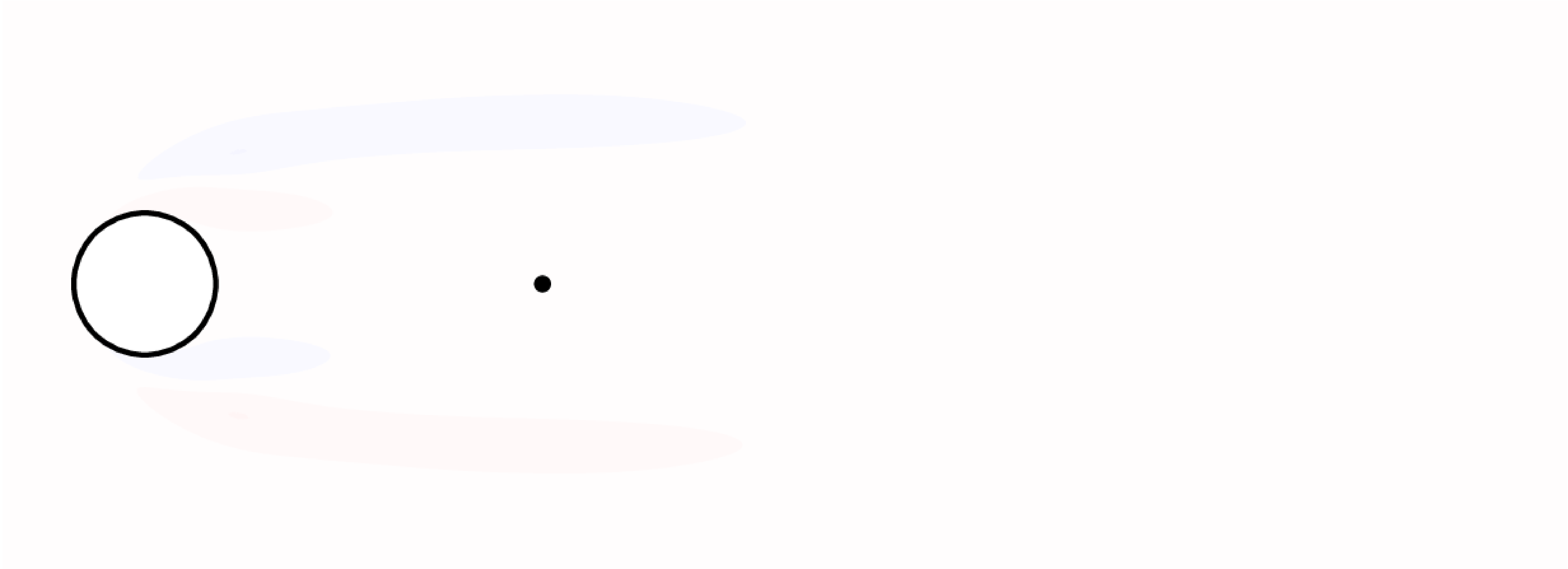}
        \includegraphics[width=1\textwidth]{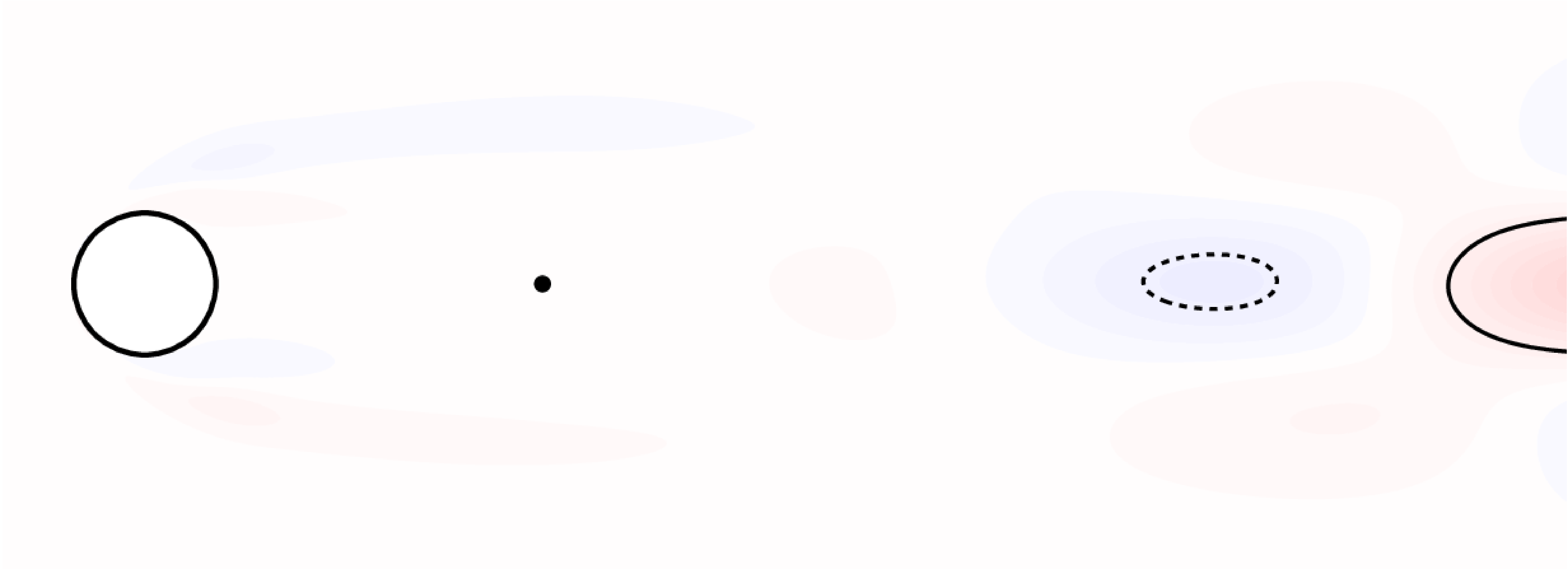}
        \includegraphics[width=1\textwidth]{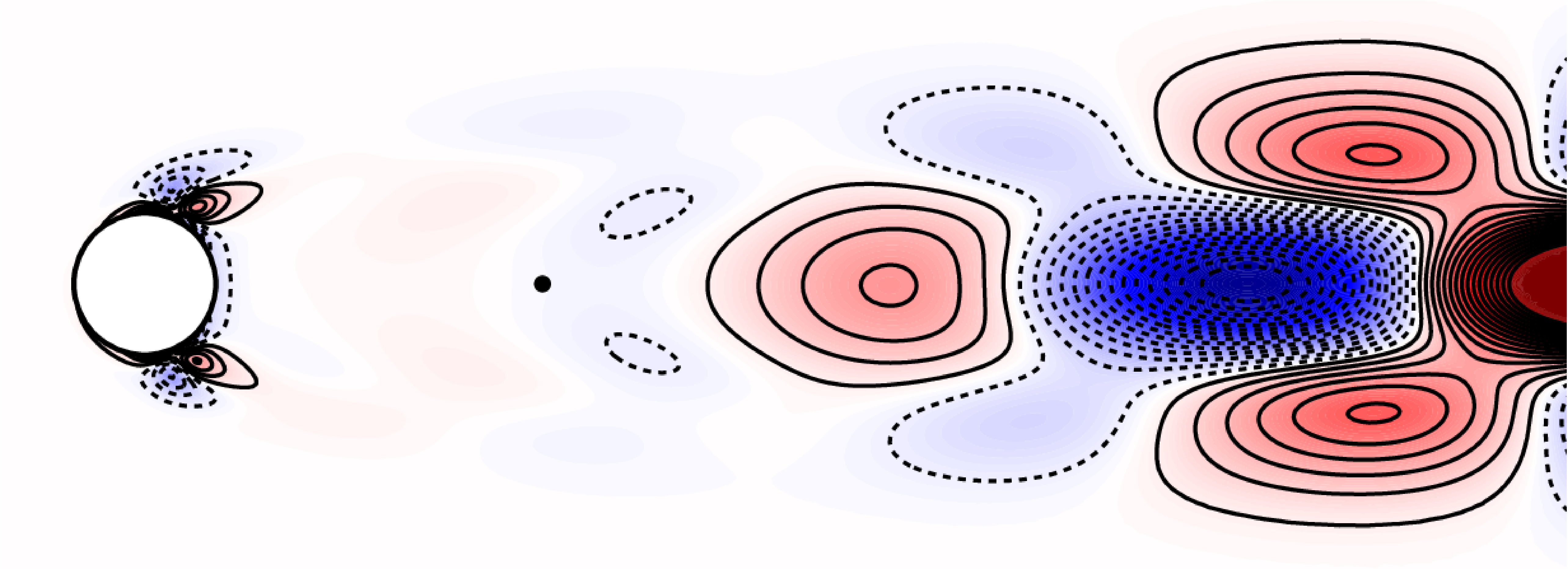}
        \begin{flushright}
        \begin{tabular}{@{}l cccccccccc@{}}
            $\ $&$\ $&$\ $& $\ $&$\ $& $\ $&$\ $& $\ $&$\ $& $\ $ &$\ $\\
            \toprule
            $\ $&$Re$&$\ $& $\textit{k}$&$\ $& $\textit{a}$&$\ $& $\textit{b}_{opt}$&$\ $& $\textit{d}$ &$\ $\\
            \midrule
            $\ $&$60$&$\ $& $2.6$&$\ $& $9.9$&$\ $& $0.3952$&$\ $& $2.8$&$\ $\\
            $\ $&$80$&$\ $& $4.6$&$\ $& $10.0$&$\ $& $0.3005$&$\ $& $2.8$&$\ $\\
            $\ $&$100$&$\ $& $11.3$&$\ $& $1.0$&$\ $& $0.2495$&$\ $& $2.8$&$\ $\\
            \bottomrule
        \end{tabular}
        \end{flushright}
        \vspace{3mm}
    \end{minipage}
    \llap{\parbox[b]{2.8in}{(a)\\\rule{0ex}{3.5in}}\parbox[b]{0in}{(b)\\\rule{0ex}{3.5in}}\parbox[b]{2.45in}{(c)\\\rule{0ex}{0.85in}}}
    \llap{\parbox[b]{0.0cm}{$\textit{v}\hspace{0.25mm}'_1$\\\rule{0ex}{7.95cm}}\parbox[b]{0.0cm}{$\textit{v}\hspace{0.25mm}'_2$\\\rule{0ex}{5.7cm}}\parbox[b]{4.6cm}{$\textit{v}\hspace{0.25mm}'_3$\\\rule{0ex}{3.45cm}}}
  \caption{DNS results of closed-loop systems. (\textit{a}) Time evolution of the transverse velocity at the sensor (\protect\blackdot) and the total perturbation energy $\textit{E}(t)$ in log scale at $\Rey=60$(\protect\bluelineshort\hspace{0.5mm}\protect\bluesmalldot\hspace{0.5mm}\protect\bluelineshort), $\Rey=80$ (\protect\redlineshort\hspace{1mm}\protect\redlineshort) and $\Rey=100$(\protect\blackline).  (\textit{b}) Vorticity contours (dashed lines for negative and solid lines for positive vorticity) for the perturbation systems at $t=75$ (\protect\mytriangle{black}) at $\Rey=60,\ 80,\ 100$ (from top to bottom). All contour plot share the same color range. (\textit{c}) Table of parameters.}
\label{fig:controled_1}
\end{figure}

The controllers are implemented in the full non-linear Navier-Stokes system with an impulse from the actuator of magnitude $10^{-3}$. The resulting performance of the controllers for each of the cases shown in $\S$\ref{sec:case1} is shown in figure \ref{fig:controled_1}. 

The stabilising effect of the controller for each case is observed from the time evolution of the transverse velocity ($\textit{v}'_1/\textit{v}'_2/\textit{v}'_3$) at the sensor, as shown in figure \ref{fig:controled_1}(\textit{a}). More convincing evidence for the complete suppression of vortex shedding is the total perturbation energy plotted beneath, which is defined as 
\begin{equation}
    \textit{E}(t)=\dfrac{1}{2}\int_{\Omega}(\textit{u}'^2(\textit{x},\ \textit{y},\ \textit{t})+\textit{v}'^2(\textit{x},\ \textit{y},\ \textit{t}))\ dx\hspace{0.5mm}dy\ ,
\end{equation}
where $\textit{u}'$, $\textit{v}'$ are the streamwise and transverse perturbation velocity components, respectively. It is clear that the controller for $\Rey=60$ performs best with the strongest attenuation. Although the perturbation energy for $\Rey=100$ remains large after $t=150$, a decreasing trend is quite clear. Figure \ref{fig:controled_1}($\textit{b}$) shows the instantaneous vorticity perturbation fields for all three cases at $t=75$. The perturbation vorticity at $\Rey=60$ stays small without observable vortex shedding whereas at a higher Reynolds number of $80$, the vorticity remains small near the sensor but is stronger downstream. At the highest Reynolds number of $100$, the vorticity is far stronger with clear vortex shedding, which is consistent with the oscillations of $\textit{v}'_3$. From the comparison between all three cases, we can see that perturbations are harder to control at higher Reynolds numbers.

The performance of each controller can be quantified using the optimal stability margin $\textit{b}_{opt}$, which is summarised in figure \ref{fig:controled_1}($\textit{c}$) together with other important parameters. Based on the comparison of $\textit{b}_{opt}$ between the three cases, we can draw a conclusion consistent with the analysis above: the higher the Reynolds number, the smaller the optimal stability margin $\textit{b}_{opt}$ the optimal feedback controller achieves.

\subsection{Optimal sensor placements}
\begin{figure}
 \vspace{2mm}
    \centerline{\includegraphics[width=0.975\textwidth]{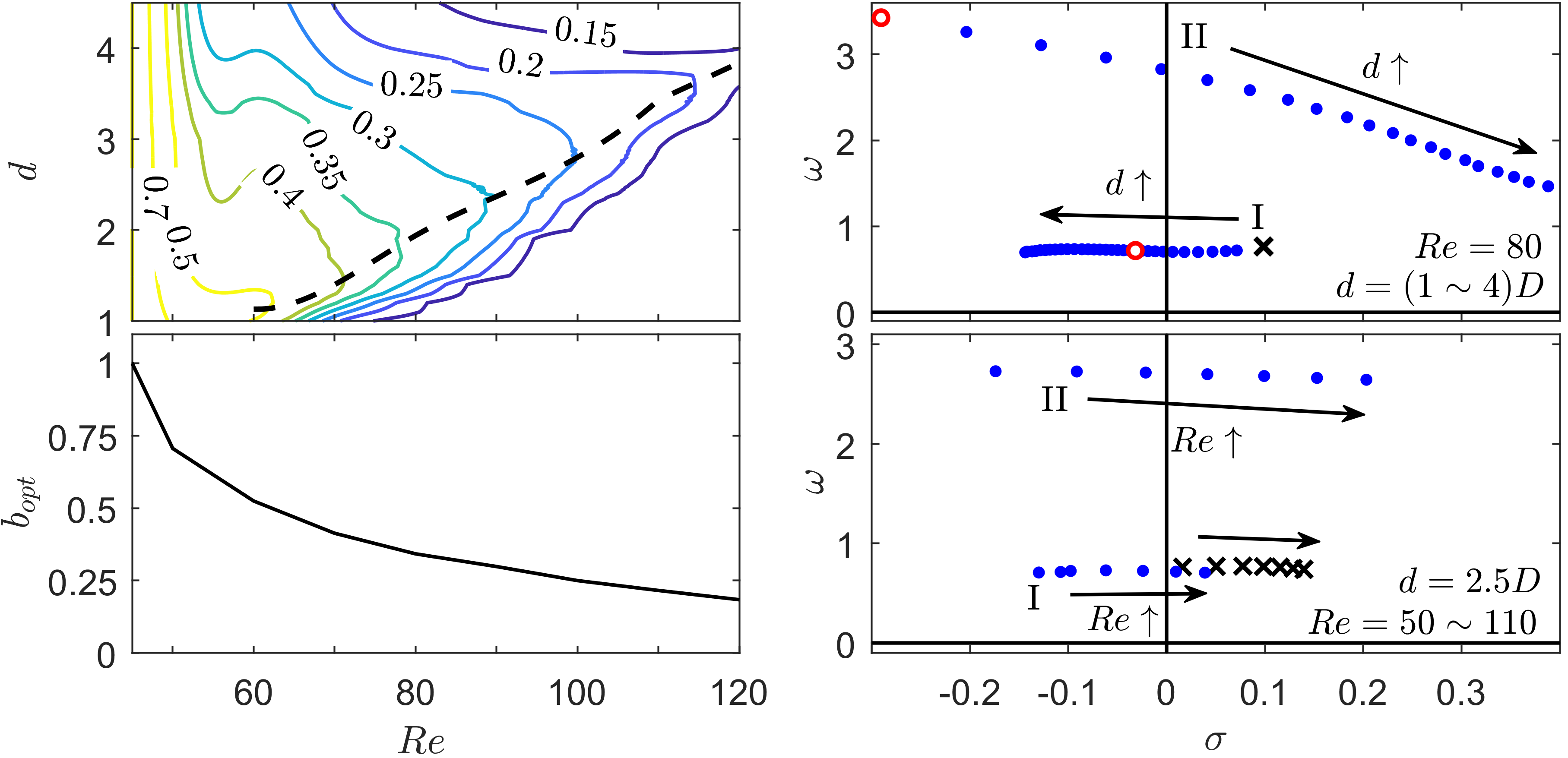}
    \llap{\parbox[b]{5.2in}{(a)\\\rule{0ex}{2.475in}}}
    \llap{\parbox[b]{2.8in}{(b)\\\rule{0ex}{2.475in}}}}
    \caption{(\textit{a}) Optimal sensor locations (the ridge  \protect\blacklineshort\hspace{1mm}\protect\blacklineshort) and contour plot of optimal stability margin $\textit{b}_{opt}$ against Reynolds number and sensor location $\textit{d}$. And the largest  $\textit{b}_{opt}$ (\protect\blackline) can be achieved at different Reynolds numbers is plotted beneath. (\textit{b}) Loci of unstable poles (\protect\xmark) and critical zeros (\protect\bluedot/\protect\redcir) of transfer functions $\widetilde{P}(s)$ for two cases. Top: different sensor locations at $\Rey=80$. Bottom: different Reynolds numbers with a sensor placed at $d=2.5D$ downstream the cylinder.}
    \label{fig:optimalplace}
\end{figure}

The preliminary investigation summarised in figure \ref{fig:controled_1} indicates a severe deterioration of control performance with increasing Reynolds number for a fixed velocity sensor.  To draw more general conclusions about control performance, we vary the position of the velocity sensor along the centreline at different Reynolds numbers. For each case, the position and the form of the actuator are unchanged, and we identify a new reduced-order model from response data at the corresponding Reynolds number. The performance of the $\mathcal{H}_{\infty}$-optimal controller for each case is quantified by the optimal stability margin $\textit{b}_{opt}$ and summarised as a function of Reynolds number and sensor location $\textit{d}$ in figure \ref{fig:optimalplace}(\textit{a}). 

First, we focus on the optimal sensor location where the optimal controller shows the best performance at each Reynolds number. Generally, the ideal position for a sensor should allow not only the measurement of the instability developing downstream but also the timely feedback of information to the actuator. Figure \ref{fig:optimalplace}(\textit{a}) first shows a contour map of the optimal stability margin $\textit{b}_{opt}$ against Reynolds number and sensor location $\textit{d}$. It is clear that a ridge exists which indicates the optimal sensor location as a function of Reynolds number. At each Reynolds number, the optimal controller for the sensor at the ridge line performs better than those for other sensor locations. However, the optimal stability margin $\textit{b}_{opt}$ on the ridge decreases sharply with increasing Reynolds number, which is shown in the figure beneath the contour map. 

Therefore, a fundamental trade-off can be concluded from figure \ref{fig:optimalplace}(\textit{a}): the sensor should be close enough to the cylinder to reduce the time delay due to convection, but it should also be far enough from the cylinder to measure important information (e.g. unstable eigenmodes) developing downstream. The compromise between these two conflicting requirements becomes harder to satisfy with increasing Reynolds number, which leads to the optimal sensor location moving downstream linearly. Similar results have been observed in recent work \citep{oehler2018sensor} that considers feedback control of the linearised Ginzburg-Landau system.

Generally, the performance and robustness  of an optimal controller, as quantified by the optimal stability margin $\textit{b}_{opt}$, can be linked to the zeros and poles of the corresponding system. Mathematically, poles and zeros of a system are roots of the denominator and numerator of the corresponding transfer function, which determine whether the system is stable, and how the system performs. More specifically, poles capture the form of each component in the system response, whereas zeros reflect how these components combine together, including the phase and magnitude of each component generated by each pole. To investigate these roots in the perturbation system, we consider two cases: $\textrm{i})$ fixing the Reynolds number at $\textit{Re}=80$ and moving the sensor  from $\textit{d}=1$ to $\textit{d}=4$; $\textrm{ii})$ fixing the sensor position at $\textit{d}=2.5$ and increasing the Reynolds number from $50$ to $110$. The root loci of these two cases are computed, where unstable poles and two kinds of critical zeros (which we label $\textrm{I}$ and $\textrm{II}$) are identified and plotted in figure \ref{fig:optimalplace}(\textit{b}). 

In the first case (top panel in figure \ref{fig:optimalplace}(\textit{b})), the sensor is moved away from the cylinder, thus measuring information further downstream. This results in the critical zero $\textrm{I}$ moving from the right-half plane (RHP) into the left-half plane (LHP) and the critical zero $\textrm{II}$ moving from the LHP into the RHP. The existence of RHP zeros is problematic for control design because they limit the maximum bandwidth or the maximum frequency that can be controlled with good performance and robustness, as described by \citet{zhou1996robust} and \citet{hoagg2007nonminimum}. In this case, zero $\textrm{I}$ stays in the LHP for $\textit{d}>1.6\textit{D}$ whereas zero $\textrm{II}$ stays in the LHP for $\textit{d}<2.4\textit{D}$ . The optimal sensor location at the Reynolds number considered is $d\approx2.0\textit{D}$, for which all zeros stay in the LHP (as indicated by \protect\redcir\ in figure \ref{fig:optimalplace}). This is also consistent with the fundamental trade-off described above.

Increasing Reynolds number also moves the zeros into the RHP, which can be seen from the root loci plotted beneath. In this case, we fix the sensor location at $\textit{d}=2.5$ and increase Reynolds number from $60$ to $110$ in intervals of $10$. It is interesting to note that increasing Reynolds number not only increases the real part of the unstable pole but also moves these critical zeros towards the right half plane. Based on the root loci in the two cases, we can conclude that it becomes harder to find a good sensor location (where no RHP zeros occur) at higher Reynolds numbers. This difficulty leads to a degradation in the performance and robustness of the optimal controllers at higher Reynolds numbers, as depicted in figure \ref{fig:optimalplace}(\textit{a}). Furthermore, zero $\textrm{I}$ appears near the unstable pole and moves closer to the unstable pole at higher Reynolds numbers (or more upstream sensor locations), whereas zero $\textrm{II}$ remains at higher frequencies. Due to the bandwidth limitation from the RHP zeros, the optimal control design algorithm would be able to compute a better controller if zero $\textrm{I}$  stays in the LHP. Thus, the systems would prefer a sensor placed further downstream to prevent RHP zero $\textrm{I}$. This preference is shown by the contour map in figure \ref{fig:optimalplace}(\textit{a}), where a gentle slope of $\textit{b}_{opt}$ occurs if the sensor is placed downstream of the optimal location but a rapid drop occurs if the sensor is placed upstream.

Similar maps of system roots are also summarised and analysed in the work by \citet{belson2013feedback}, where optimal controllers were designed for a linearised 2D Blasius boundary layer controlled by different types and positions of actuators and sensors. A degradation of  the controllers' performance and robustness was observed when RHP zeros occurred. Generally, the physical mechanisms behind the RHP zeros are due to $\textrm{i}$) the time delay or $\textrm{ii}$) the observability of the structures that are to be controlled. In a flow system, when the sensor is far downstream of the actuator, it measures the effect of the actuator with a time delay due to the convective nature of the flow. That is, the sensor measures flow structures which convected past the actuator at an earlier time. With outdated information, the controller poorly estimates and controls the flow structures near the actuator. This time delay becomes more significant as the sensor moves downstream and results in RHP zeros in the reduced-order transfer function $\widetilde{P}(s)$. 

However, with a sensor close to the actuator, the performance and robustness of the optimal controller is still restricted by RHP zeros that occur near the unstable pole. Such RHP zeros cancel the effect of the unstable pole and prevent the sensor from measuring the instability. In other words, the poor performance and robustness of the controller is caused by a lack of observability of the unstable mode instead of excessive time delay.

In this section, we have shown that different sensor locations and Reynolds numbers have similar properties that restrict the performance and robustness of the optimal controllers. At higher Reynolds numbers, even an optimal controller performs poorly for both control setups. In other words, the best possible performance that can be achieved is severely restricted. From the perspective of control theory, we observe RHP zeros which limit control performance. Thus, we cannot always find a controller with good performance and robustness, and this can be attributed to the compromise between the observability of the instability and the size of the convective time delay. 

\begin{figure}
 \vspace{2mm}
    \centerline{
    \hspace{1mm}
    \includegraphics[width=0.3\textwidth]{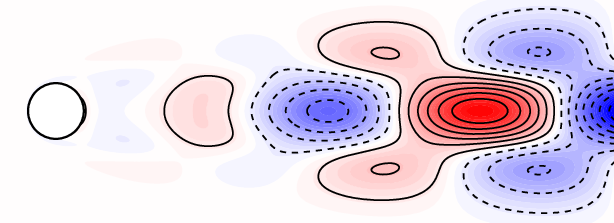}
    \llap{\parbox[b]{1.8in}{(a)\\\rule{0ex}{0.575in}}}
    \hspace{1mm}
    \includegraphics[width=0.3\textwidth]{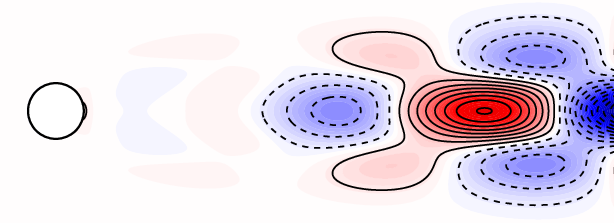}
    \llap{\parbox[b]{1.8in}{(b)\\\rule{0ex}{0.575in}}}
    \hspace{1mm}
    \includegraphics[width=0.3\textwidth]{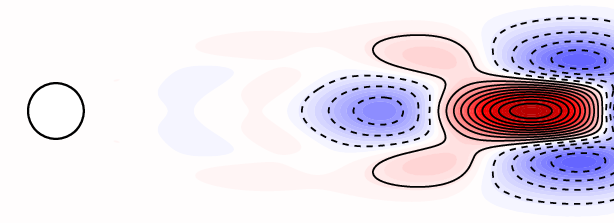}
    \llap{\parbox[b]{1.8in}{(c)\\\rule{0ex}{0.575in}}}
    }
    \caption{Vorticity contours (dashed lines for negative and solid lines for positive vorticity) for the perturbation system (normalised, real part) actuated by the harmonic oscillation of the cylinder at the unstable frequency for (\textit{a}) $\Rey=60$, (\textit{b}) $\Rey=80$, (\textit{c}) $\Rey=100$. Contour plots share the same scale.}
    \label{fig:osci_response}
\end{figure}
\section{Body-mounted control setup case}\label{sec:bodyset}
We now turn our attention to a more physically representative control setup with a body-mounted actuator and a body-mounted sensor. The schematic diagram is illustrated in figure \ref{fig:controlsetups}(\textit{b}), where the flow field is now controlled by the oscillation of the cylinder itself, which oscillates in response to the lift measured on the cylinder. Following a similar procedure as the in-flow control case, we also consider the physics behind the difficulty in synthesizing controllers with good performance and robustness.
\subsection{Open-loop system identification}
The purpose of both control setups is to eliminate the perturbations and drive the system towards the steady solution: the base flow. Figure \ref{fig:osci_response} shows the normalised vorticity field of the corresponding perturbation system $P(s)$ actuated by the oscillation of the cylinder at the instability frequency (i.e. the resolvent operator between the input and output in \eqref{equ:pertur_sys}). Similar to the in-flow setup, the large vortical structure actuated on by the moving cylinder develops further downstream as the Reynolds number increases.

\begin{figure}
 \vspace{2mm}
 \hspace{0.5mm}
    \centerline{\includegraphics[width=0.95\textwidth]{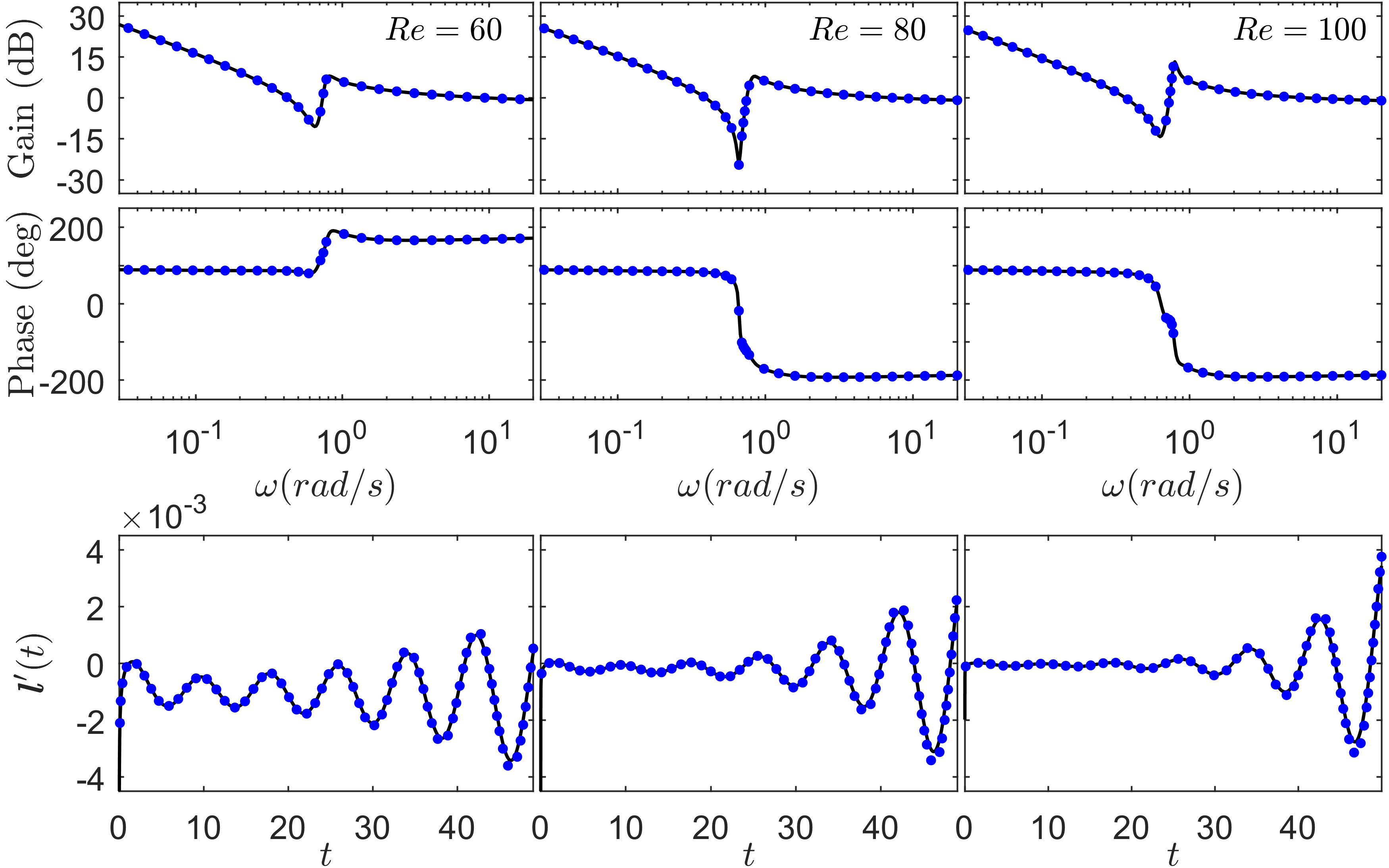}
    \llap{\parbox[b]{5.2in}{(a)\\\rule{0ex}{3.1in}}}
    \llap{\parbox[b]{5.25in}{(b)\\\rule{0ex}{1.175in}}}}
    \caption{(\textit{a}) Frequency responses from $P(j\omega)$ (\protect\blackline) compared to those from reduced-order models $\widetilde{P}(j\omega)$ (\protect\bluedot) at $\Rey=60,\ 80,\ 100$. (\textit{b}) The corresponding open-loop impulse responses from numerical simulations. The results for $\Rey=60,\ 80$ are multiplied by 15 and 3, respectively, so that the same scale can be used.}
    \label{fig:modelvalidation_2}
\end{figure}

The system identification procedure is carried out in a similar manner to \S \ref{sec:case1} and summarised in figure \ref{fig:modelvalidation_2}. The reduced-order models $\widetilde{P}(s)$ are chosen such that the fitting residual $\epsilon$ is below $10^{-5}$ with orders less than 35. The Bode plots of identified transfer functions between the actuator and the sensor are shown in figure \ref{fig:modelvalidation_2}(\textit{a}) and compared to the frequency responses of the true systems. Unlike the in-flow setup, the perturbation system with body-mounted setup has infinite zero-frequency response (i.e. the system contains an integrator) and constant infinite-frequency response (i.e. the system contains a non-zero feed-forward term). 

Figure \ref{fig:modelvalidation_2}(\textit{b}) shows comparisons of open-loop impulse responses (of magnitude $10^{-4}$) from the identified models and direct numerical simulations (DNS). The excellent agreement observed validates the accuracy of the reduced-order models.

\subsection{Model-based feedback control}
\begin{figure}
\vspace{1.5mm}
\hspace{1.75mm}
    \begin{minipage}[b]{0.5\textwidth}
    \centering
    \includegraphics[width=1\textwidth]{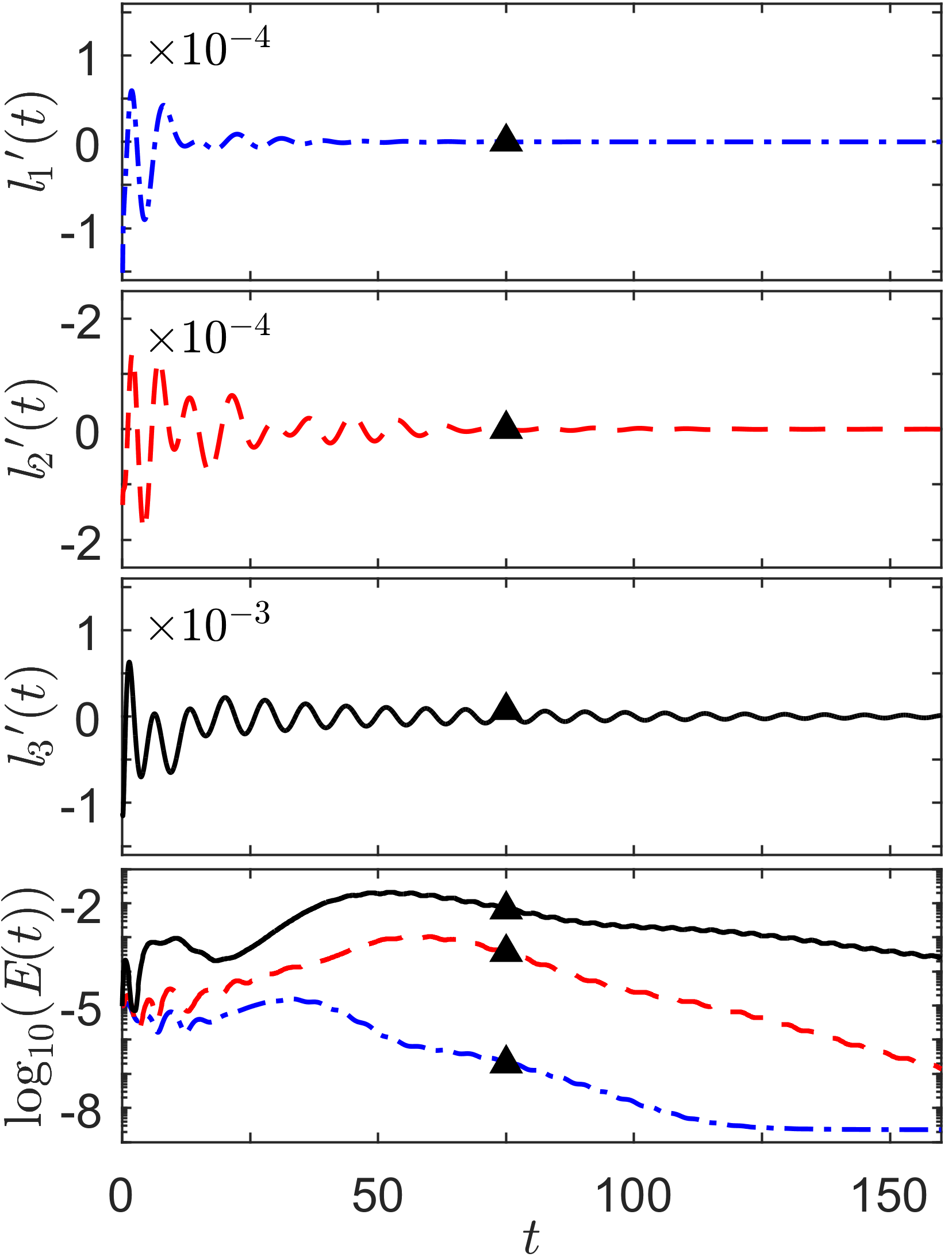}
    \end{minipage}
    \begin{minipage}[b]{0.45\textwidth}
        \centering
        \includegraphics[width=1\textwidth]{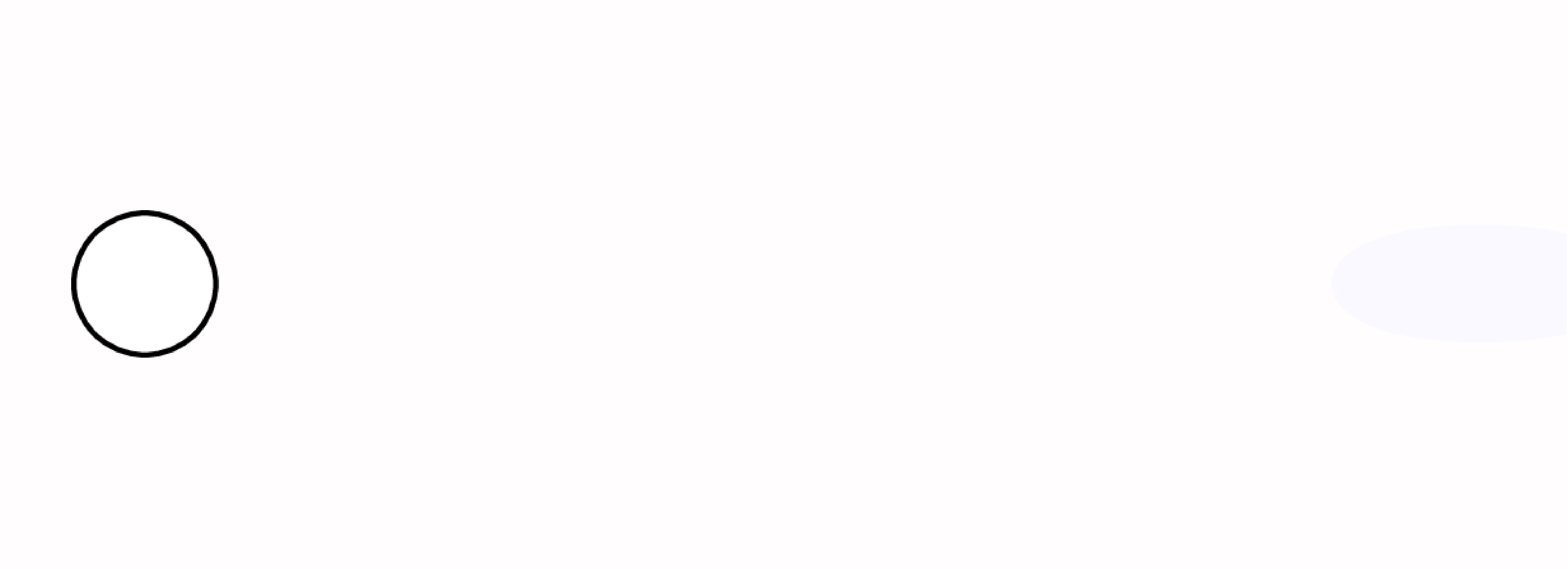}
        \includegraphics[width=1\textwidth]{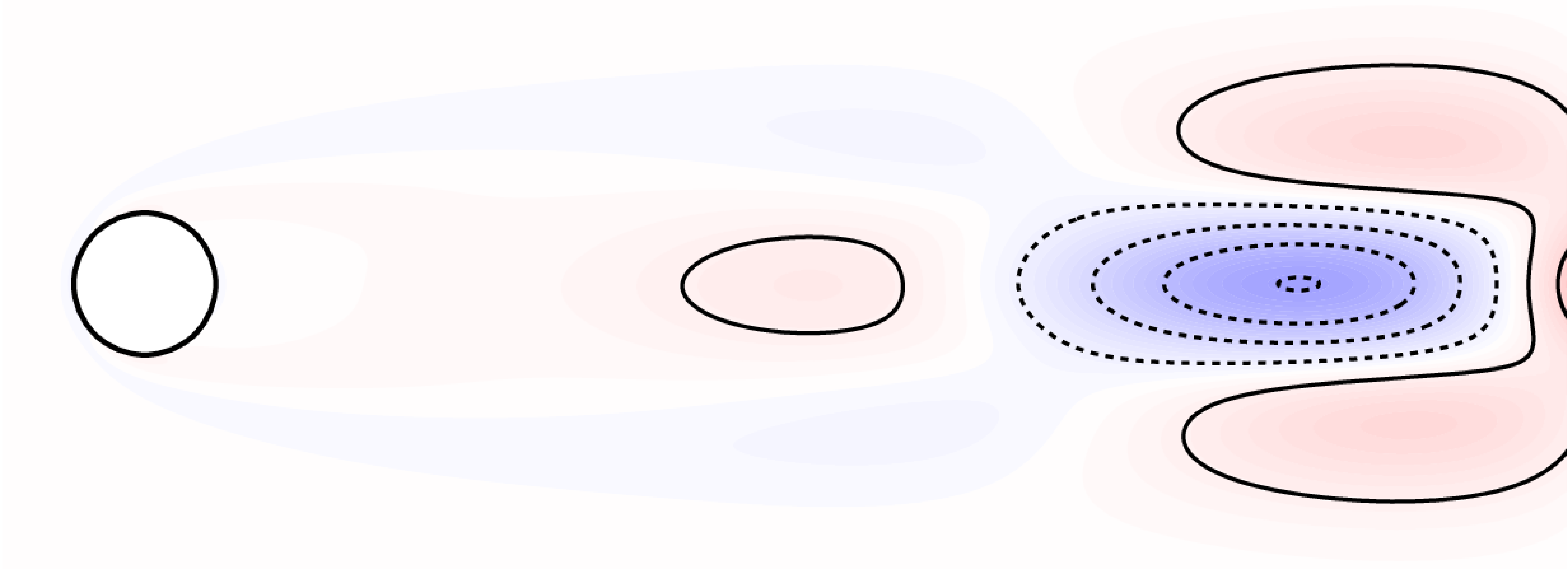}
        \includegraphics[width=1\textwidth]{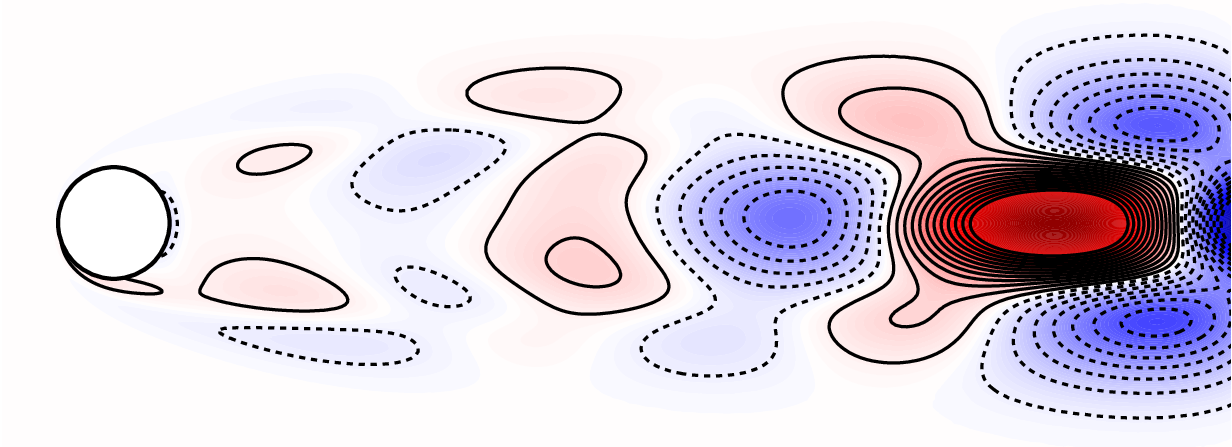}
        \begin{flushright}
        \begin{tabular}{@{}l cccccccccc@{}}
            $\ $&$\ $&$\ $& $\ $&$\ $& $\ $&$\ $& $\ $&$\ $& $\ $ &$\ $\\
            \toprule
            $\ $&$Re$&$\ $& $\textit{k}$&$\ $& $\textit{a}$&$\ $& $\textit{b}_{opt}$&$\ $& $\textit{d}$ &$\ $\\
            \midrule
            $\ $&$60$&$\ $& $3.1$&$\ $& $0.5$&$\ $& $0.2537$&$\ $& $--$&$\ $\\
            $\ $&$80$&$\ $& $20.0$&$\ $& $0.2$&$\ $& $0.0764$&$\ $& $--$&$\ $\\
            $\ $&$100$&$\ $& $0.7$&$\ $& $19.8$&$\ $& $0.0313$&$\ $& $--$&$\ $\\
            \bottomrule
        \end{tabular}
        \end{flushright}
        \vspace{3mm}
    \end{minipage}
    \llap{\parbox[b]{2.65in}{(a)\\\rule{0ex}{3.5in}}\parbox[b]{0in}{(b)\\\rule{0ex}{3.5in}}\parbox[b]{2.45in}{(c)\\\rule{0ex}{0.85in}}}
  \caption{DNS results of closed-loop systems. (\textit{a}) Time evolution of the cylinder lift and the total perturbation energy $\textit{E}(t)$ in log scale at $\Rey=60$(\protect\bluelineshort\hspace{0.5mm}\protect\bluesmalldot\hspace{0.5mm}\protect\bluelineshort), $\Rey=80$ (\protect\redlineshort\hspace{1mm}\protect\redlineshort) and $\Rey=100$(\protect\blackline).  (\textit{b}) Vorticity contours (dashed lines for negative and solid lines for positive vorticity) for the perturbation systems at $t=75$ (\protect\mytriangle{black}) at $\Rey=60,\ 80,\ 100$ (from top to bottom). All contour plot share the same color range. (\textit{c}) Table of parameters.}
\label{fig:controled_2}
\end{figure}

Following the same procedure as in $\S$\ref{sec:feedcon1}, we design optimal controllers for reduced-order models and implement them in the full non-linear Navier-Stokes system actuated by an initial impulse of magnitude $10^{-4}$. The parameters of controllers and the corresponding closed-loop simulations are summarised in figure \ref{fig:controled_2}. The stabilisation of vortex shedding is achieved only up to $\Rey=100$, which can be seen both in the time evolution of the lift $l\hspace{0.1mm}'(t)$ and in the total perturbation energy $E(t)$. The comparison among simulations at three Reynolds numbers indicates a similar deterioration of control performance to that seen for the in-flow setup of $\S 4$.

The control of vortex shedding using such body-mounted setup is more challenging than control with the in-flow setup in $\S$\ref{sec:in-flow_case}. This is revealed by closed-loop simulations in three ways. First, controllers designed for $\Rey=60,\ 80,\ 100$, although stabilising, show poorer performance than controllers designed for the in-flow setup. Second, the optimal controller fails to stabilise the flow system if the Reynolds number is greater than 100. Third, the optimal stability margin $b_{opt}$, which is a performance indicator of the controller, decreases from $0.2537$ at $\Rey=60$ to an extremely small value of $0.0313$ at $\Rey=100$. This is a much more severe degradation than that seen for the in-flow setup which decreased from $0.3952$ at $\Rey=60$ to $0.2495$ at $\Rey=100$.

\begin{figure}
 \vspace{2mm}
    \centerline{
    \hspace{1mm}
    \includegraphics[width=0.95\textwidth]{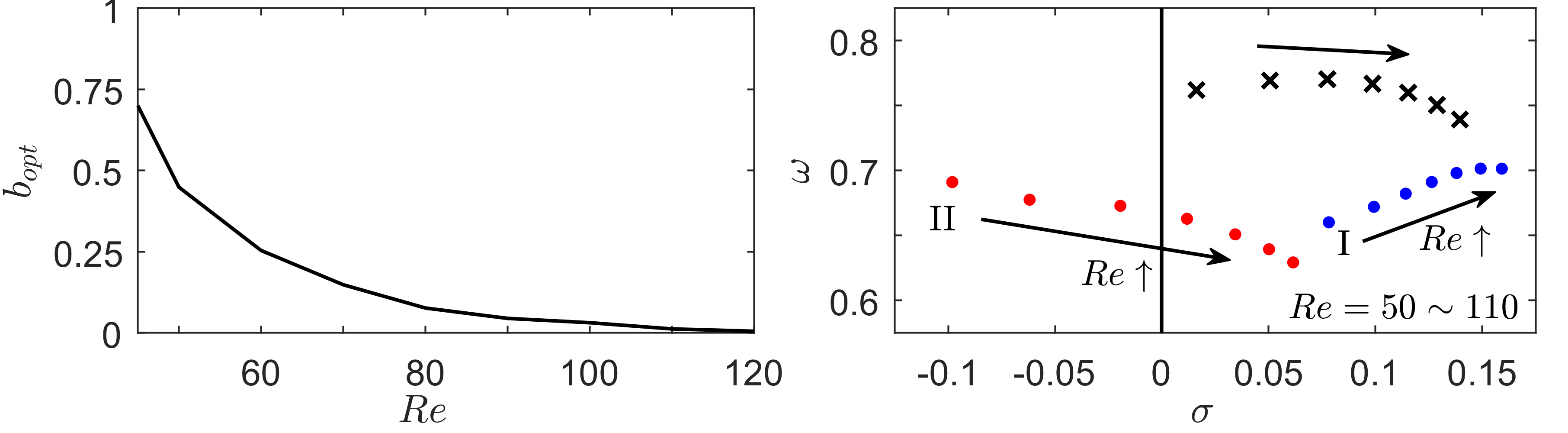}
    \llap{\parbox[b]{5.2in}{(a)\\\rule{0ex}{1.35in}}}
    \llap{\parbox[b]{2.65in}{(b)\\\rule{0ex}{1.35in}}}
    }
    \vspace{0.5mm}
    \centerline{
    \hspace{1mm}
    \includegraphics[width=0.95\textwidth]{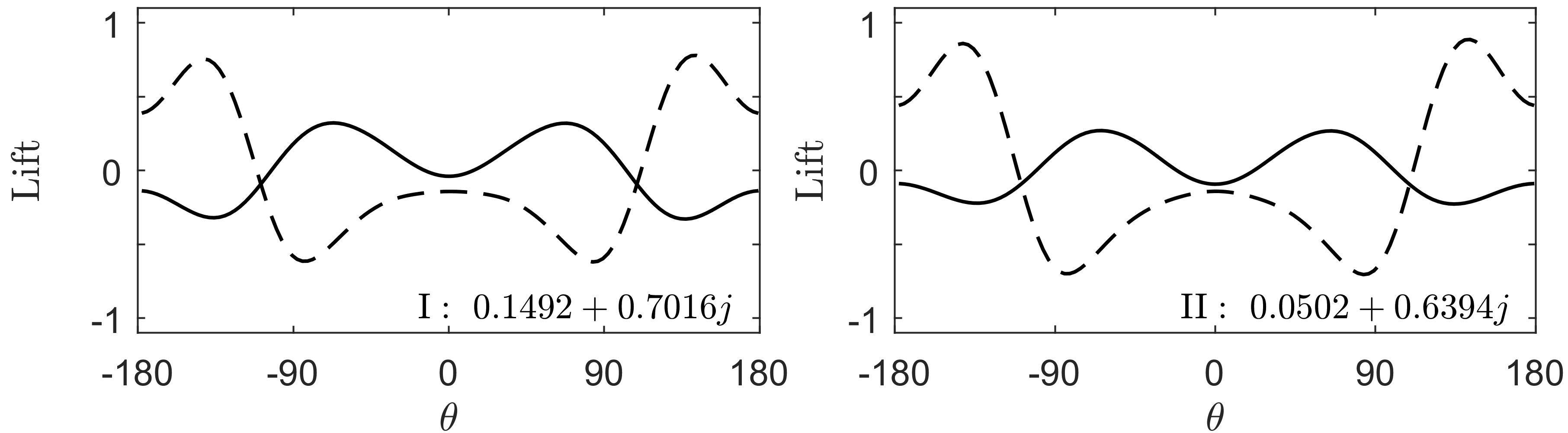}
    \llap{\parbox[b]{5.2in}{(c)\\\rule{0ex}{1.35in}}}
    }
    \caption{(\textit{a}) The largest  $\textit{b}_{opt}$ (\protect\blackline) that can be achieved at different Reynolds numbers. (\textit{b}) Loci of unstable poles (\protect\xmark) and critical zeros (\protect\bluedot/\protect\reddot) of transfer functions $\widetilde{P}(s)$ at different Reynolds numbers. (\textit{c}) Lift distributions (\protect\blackline\ for real part and \protect\blacklineshort\hspace{1mm}\protect\blacklineshort\ for imaginary part) on the cylinder at RHP zeros $\textrm{I}/\textrm{II}$ and $\Rey=100$.}
    \label{fig:control2}
\end{figure}

A more detailed trend of the optimal stability margin $b_{opt}$ is depicted in figure \ref{fig:control2}(\textit{a}) as a function of Reynolds number. The severe degradation of control performance is clearly shown by the reduction in $b_{opt}$ from $0.7$ at $\Rey=45$ to $0.0313$ at $\Rey=100$, whereas the optimal stability margin $b_{opt}$ of the in-flow setup changes from $1.0$ to $0.2495$ in the same range of Reynolds numbers. 

From the perspective of control theory, the performance and robustness of an optimal controller, as quantified by the optimal stability margin $b_{opt}$, can be affected by the roots (zeros and poles) of the corresponding system, especially those near the unstable mode. Figure \ref{fig:control2}(\textit{b}) shows the root loci near the unstable mode for the flow systems between $\Rey=50$ and $\Rey=110$. It can be seen from the figure that as Reynolds number increases, these poles and zeros move into the right-half plane (RHP), which implies stronger instability and RHP zeros. More importantly, the RHP zero $\textrm{I}$  moves closer to the unstable poles at higher Reynolds numbers, which cancels the effect of instability and reduces the observability of the unstable mode. RHP zeros are problematic for control design and restrict the performance and robustness of optimal controllers. Similar restrictions are also observed from the root loci of the flow system with an in-flow setup. However, at low Reynolds numbers, the in-flow setup does not have RHP zeros, whereas in the system with body-mounted setup, at least one RHP zero occurs.  

For the body-mounted setup, the sensor and actuator are attached to the cylinder and no time delay should be present in the transfer function from the actuator to the sensor. Therefore to investigate the physical mechanisms represented by these zeros, we take the cylinder flow at $\Rey=100$ as an example and plot the lift distributions on the cylinder surface for two RHP zeros in figure \ref{fig:control2}(\textit{c}). It is interesting to note that even though zero lift should act on the cylinder at these zeros, the force acting on the cylinder surface is nonzero. Instead, the vertical force acting on the rear surface ($-90^{\circ}<\theta<90^{\circ}$) cancels the force acting on the front surface. Such force cancellations always occur near the unstable pole which prevent the lift sensor from measuring the instability developed in the system. 

\section{Conclusions}\label{sec:cons}
Model-based feedback control of vortex shedding behind a 2D circular cylinder has been investigated. We have built linear reduced-order models for the linear dynamics of the flow systems about unstable equilibrium states in an input-output framework. The method involves the formulation of the resolvent operator from the linearised flow system to compute frequency responses for control setups. We then identify low-order models from these responses using vector fitting methods. The resulting models are capable of accurately capturing the input-output flow behaviour which has been demonstrated by comparing their impulse responses to DNS. Stabilising feedback controllers have been designed for these ROMs based on the $\mathcal{H}_{\infty}$ loop-shaping procedure of \citet{glover1989robust}, and validated by applying them to the original high-order nonlinear flow systems. The method has been applied to two single-input single-output control configurations and led to the full stabilisation of the nonlinear flow systems over a range of Reynolds numbers.

To investigate the role of sensor placement on the performance and robustness of the closed-loop, we chose an in-flow control setup with a pair of anti-symmetrical body forces for actuation and a downstream point sensor measuring the vertical velocity. Different sensor positions have been examined for a range of Reynolds numbers and the corresponding optimal sensor locations indicate two principles of the sensor placement: on the one hand, it should be close enough to the actuator to reduce the time delay due to highly convective nature of the system; on the other hand, the sensor is required to measure the information developing downstream. 

Similar trade-offs have also been observed in the feedback control of the linearised Ginzburg-Landau system by \citet{oehler2018sensor}. Difficulties in satisfying these two conflicting requirements at a higher Reynolds number causes a deterioration in the optimal controller's performance and robustness. This deterioration has been shown to be related to the presence of RHP zeros in the transfer function from actuator to sensor as in \citet{belson2013feedback},  and the sensor location has been observed to be a significant parameter in feedback flow control. These RHP zeros prevent robust control tools from designing well-behaved controllers and place severe limitations on closed-loop performance.

The method has also been applied to a body-mounted control setup in which the controller measures the lift on the cylinder and actuates the cylinder's displacement. We controlled the acceleration of the oscillatory cylinder according to its lift measurement and successfully suppressed vortex shedding up to \Rey=100. The deterioration of the optimal controller's performance and robustness is more severe than for the in-flow sensor setup at all Reynolds numbers considered. In this control setup both the sensor and actuator are attached to the cylinder, which implies zero time delay, but the instability developing downstream cannot be sufficiently well captured. This limitation manifests itself in the form of lift cancellations which are represented by RHP zeros near the unstable poles.

This study has presented an efficient modelling approach using the resolvent operator, which does not rely either on expensive direct numerical simulations or experimental data. It constructs models of sufficient quality to design and implement robust stabilising controllers for unstable flows. In addition, the formulation accounts for actuation and measurements directly, and therefore, it can be applied to a broad range of flow control configurations. 

\newpage
\bibliographystyle{jfm}
\bibliography{jfm-instructions}

\end{document}